\documentclass[fleqn,usenatbib]{mnras} 
\usepackage{ae,aecompl}
\usepackage{ulem}
\usepackage{xargs} % Use more than one optional parameter in a new commands
\usepackage{siunitx}
\usepackage{longtable}
\usepackage{booktabs}
\usepackage{pdflscape}

\usepackage{amsmath}	% Advanced maths commands
\usepackage{natbib}
\usepackage{color}
\usepackage{lscape}
\usepackage{graphicx} 
\usepackage{amssymb}

\newcommand{\geqsim}{\,\raisebox{-0.6ex}{$\buildrel > \over \sim$}\,}

\newcommand{\teff}{$T_{\rm eff}$}
\newcommand{\logg}{$\log g$}
\newcommand{\vsini}{$v\sin i$}

\newcommand{\COMPO}{{\sf COMPO2}}
\newcommand{\kms}{km\,s$^{-1}$}
 
\newcommand{\gaia}{{\it Gaia}}

\newcommand{\Msun}{{$M_{\odot}$}}
\newcommand{\Lsun}{{$L_{\odot}$}}
\newcommand{\Rsun}{{$R_{\odot}$}}

\title[New, late-type spectroscopic binaries with X-ray emission]{New, late-type spectroscopic binaries with X-ray emission\thanks{Based on observations made with the Catania Astrophysical Observatory Spectropolarimeter (CAOS) operated by the Catania Astrophysical Observatory.}} 
\author[A. Frasca et al.]{A. Frasca$^{1}$\thanks{E-mail: antonio.frasca@inaf.it}, 
G. Catanzaro$^{1}$, 
I. Bus\`a$^{1}$, 
P. Guillout$^{2}$, %A. Klutsch$^{1,2}$, 
J. Alonso-Santiago$^{1}$,
C. Ferrara$^{3,1}$, 
\and 
M. Giarrusso$^{4,1}$, M. Munari$^{1}$, F. Leone$^{3,1}$ \\
~\\  
$^{1}$INAF-Osservatorio Astrofisico di Catania, Via S.Sofia 78, I-95123, Catania, Italy\\ 
$^{2}$Universit\'e de Strasbourg, CNRS, Observatoire Astronomique de Strasbourg, UMR 7550, F-67000 Strasbourg, France\\
$^{3}$Universit\`a degli studi di Catania, Via S.Sofia 78, I-95123, Catania, Italy\\
$^{4}$Universit\`a degli Studi di Firenze, Dipartimento di Fisica e Astronomia, Largo E. Fermi 2, I-50125 Firenze, Italy
} 
 
\date{Accepted 2022 June 30. Received 2022 June 24} % in original form } 
 
\pubyear{2022}

\begin{document} 
\label{firstpage} 
\pagerange{\pageref{firstpage}--\pageref{lastpage}}

\maketitle 
 
\begin{abstract} 
In this paper we present a spectroscopic study of six double-lined binaries, five of which were recently discovered in a high-resolution
spectroscopic survey of optical counterparts of stellar X-ray sources. Thanks to high-resolution spectra acquired with CAOS spectropolarimeter during seven years, we were able to measure the radial velocities of their components and determine their orbital elements.  We have applied our code COMPO2 to determine the spectral types and atmospheric parameters of the components of these spectroscopic binaries and found that two of these systems are composed of main sequence stars, while the other four contain at least one evolved (giant or subgiant) component, similar to other well-known RS\,CVn systems. The subtraction of a photospheric template built up with spectra of non-active stars of the same spectral type as those of the components of each system has allowed us to investigate the chromospheric emission that fills in the H$\alpha$ cores.
We found that the colder component is normally the one with the largest H$\alpha$ emission. 
None of the systems show a detectable Li{\sc i}$\lambda$6708 line, with the exception of TYC\,4279-1821-1, which exhibits high photospheric abundances in both components.
Photometric time series from the literature allowed us to assess that the five systems with a nearly circular orbit have also photometric periods close or equal to the orbital ones, indicating spin-orbit synchronization. For the system with a highly eccentric orbit, a possible pseudo-synchronization with the periastron velocity is suggested.
\end{abstract} 
 

\section{Introduction}

X-ray emission is nowadays one of the best and accessible indicators of high-energy phenomena in stellar atmospheres and circumstellar environments.
It is particularly useful to detect magnetic activity in late-type (FGKM) stars. Large area surveys, such as the 
ROSAT All-Sky Survey (RASS, \citealt{Voges1999}) and the 3XMM-DR7 catalogue \citep{3XMM-DR7},  are ideal tools for the identification of
young and/or active stars.

During the course of a survey of stellar X-ray sources selected by the cross-correlation of the RASS and TYCHO \citep{HIPPA1997} catalogues (henceforth 
the {\it RasTyc} sample), \citet{Guillout2009}
and \citet{Frasca2018} discovered a number of double-lined spectroscopic binaries (SB2s) and multiple systems. Unlike single objects, which are mainly young 
main-sequence (MS) stars, the high X-ray activity level for close binaries is often caused by the tidal synchronization of orbital and rotation motions 
that forces the components of these system to spin faster, enhancing the dynamo action.  The most active binaries are those of the RS~CVn type, which 
have periods from a few days to a few tens of days and contain at least one evolved (giant or subgiant) component 
\citep[e.g.,][and references therein]{Hall1976,Eker2008}.
However, binaries composed of young MS stars, as witnessed by photospheric lithium in both components, which are likely BY\,Dra variables, are also present in X-ray selected samples.

Close binaries, especially spectrophotometric ones, are very important because they allow us to derive the basic physical properties of their components,
such as effective temperatures, radii, and masses.

In this paper we present spectroscopic follow-up observations of six double-lined binaries, five of which were recently discovered by \citet{Frasca2018} during a spectroscopic survey of {\it RasTyc} faint sources.

\section{Observation and data reduction}
\label{obs}

\setlength{\tabcolsep}{5pt}

\begin{table*}	%[ht]
\caption[SB2 sources]{Parameters of the investigated binaries from the literature.}
\begin{center}
\begin{tabular}{lccccrrrccc}
\hline
%\noalign{\smallskip}
           Name  &  RA (2000) & DEC (2000)                &  $V^{\rm a}$  & $B$--$V^{\rm a}$ & $\pi^{\rm b}$~~~~~~~~ & $\mu_{\alpha}^{\rm b}$~~~ & $\mu_{\delta}^{\rm b}$~~~ &   X-ray source &  Counts \\
                 &  h m s     & $\degr ~\arcmin ~\arcsec$ & (mag)         & (mag)            & (mas)~~~~~~~          & {\scriptsize (mas/yr)}    & {\scriptsize (mas/yr)}    &    1RXS        &  (ct/s) \\
%\noalign{\smallskip}
\hline
%\noalign{\smallskip}
 TYC\,3386-868-1  & 06 04 51.40 & +51 42 00.8 & 9.35 &       0.909               &  3.3728\,$\pm$\,0.0176 & 15.033    & $-18.929$  & {\scriptsize J060452.0+514200} & 1.51$\times 10^{-1}$ \\ % 3.34\,$\pm$\,0.25 & 14.939    & $-18.300$  
 G\,137-52        & 15 47 11.90 & +15 09 14.9 & ~9.61$^{\rm c}$ & 0.99$^{\rm c}$ & 17.9781\,$\pm$\,0.0262 & 188.888   & $-365.224$ & {\scriptsize J154712.0+150912} & 2.57$\times 10^{-1}$ \\ %14.25\,$\pm$\,2.22 & 191.02~   & $-364.46$~ 
 BD+10\,2953      & 16 05 02.23 & +10 28 54.6 & 9.21 &       0.755               &  4.4293\,$\pm$\,0.0203 & $-16.550$ & 11.350	   & {\scriptsize J160501.0+102843} & 6.00$\times 10^{-2}$ \\ % 4.14\,$\pm$\,0.27 & $-16.229$ & 11.927 
 V1079~Her        & 16 20 13.72 & +24 36 11.1 & 9.48 &       1.111               &  2.4434\,$\pm$\,0.0134 & $-5.402$  & $-3.449$   & {\scriptsize J162013.2+243606} & 2.94$\times 10^{-1}$ \\ % 2.57\,$\pm$\,0.33 & $-5.010$  & $-4.075$ 
 BD+62\,1880      & 20 58 16.40 & +63 17 38.8 & 9.75 &       0.614               &  7.9306\,$\pm$\,0.0130 & 67.637    & 84.592     & {\scriptsize J205814.0+631750} & 4.08$\times 10^{-2}$ \\ % 7.79\,$\pm$\,0.26 & 67.657    & 84.443   %  Faint source Voges+2000
 TYC\,4279-1821-1 & 23 22 40.03 & +61 13 33.3 & 9.90 &       0.781               &  3.1552\,$\pm$\,0.0113 & $-7.357$  & $-16.165$  & {\scriptsize J232241.3+611335} & 9.92$\times 10^{-2}$ \\ % 3.35\,$\pm$\,0.40 & $-8.173$  & $-16.865$  
%\noalign{\smallskip}
\hline
\end{tabular}
\end{center}
\begin{list}{}{}								       
\item[$^{\rm a}$] $V$ magnitude and $B$--$V$ color from the TYCHO catalogue \citep{HIPPA1997}.
\item[$^{\rm b}$] Parallax and proper motions from \gaia--EDR3 \citep{GaiaEDR3}.
\item[$^{\rm c}$] From \citet{Ryan1989}.
\end{list}
\label{Tab:X}
\end{table*}

Time-resolved spectroscopy was carried out at the {\it Catania Astrophysical Observatory Spectropolarimeter} (CAOS) 
which is a fiber fed, high-resolution, cross-dispersed \'echelle spectrograph 
\citep{Catanzaro2015,Leone2016}  installed at the Cassegrain focus of the
91\,cm telescope of the {\it ``M. G. Fracastoro''} observing station of the 
Catania Astrophysical Observatory (Mt. Etna, Italy). A number of  spectra were obtained from 2015 to 2021 %in 2015 and in 2016 with 
with exposure times ranging from 1200 to 2400~sec. For the best exposed spectra the signal-to-noise ratio (SNR) was at least 50 in the 
continuum in the 4300--7000~{\AA} wavelength range. The spectral resolution of CAOS is $R\simeq$\,45\,000, as we have verified from the 
%measure of the 
full width at half maximum (FWHM) of the emission lines of the Th-Ar calibration lamp or the telluric absorption lines. 

The reduction of spectra, which included the subtraction of the bias frame, trimming, correcting for the flat-field and 
the scattered light, extraction of the orders, and wavelength calibration, was done using the NOAO/IRAF packages\footnote{IRAF 
is distributed by the National Optical Astronomy Observatory, which is operated by the Association of Universities for Research 
in Astronomy, Inc.}. 

%The IRAF package {\it rvcorrect} was used to determine the heliocentric velocity correcting the spectra for the Earth's motion. 

%We collected radial velocities for the six investigated binaries with other two instruments, namely SARG@TNG and AURELIE@OHP  
%\citep[see][for details]{Frasca2018}.

We used also radial velocities previously collected by us with other two instruments, namely SARG at the 3.5 m TNG telescope (Observatorio del Roque de los Muchachos, Canary
Islands, Spain) and AURELIE at the 1.52 m
telescope at the Observatoire de Haute Provence (OHP, France).
These data, which allowed us to classify these objects as SB2s, were published in \citet[][]{Frasca2018}.

Table~\ref{Tab:X} displays the observed targets along with  photometric and astrometric data from the literature.

\section{Data analysis and results}
\label{sec:anal}

\begin{figure*}	%[ht]
  \begin{center}
  \includegraphics[width=5.5cm,height=5.5cm]{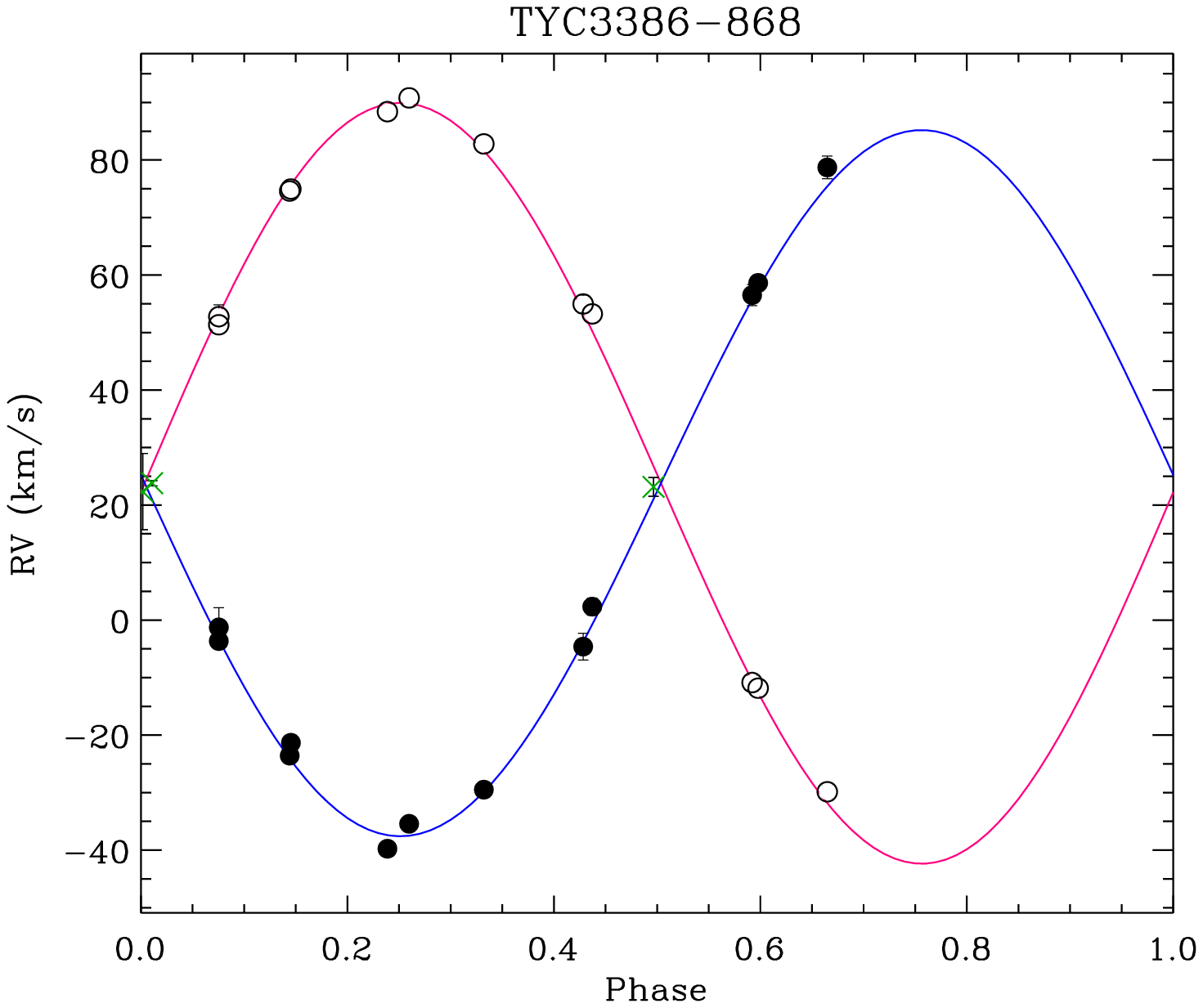}	 
  \includegraphics[width=5.5cm,height=5.5cm]{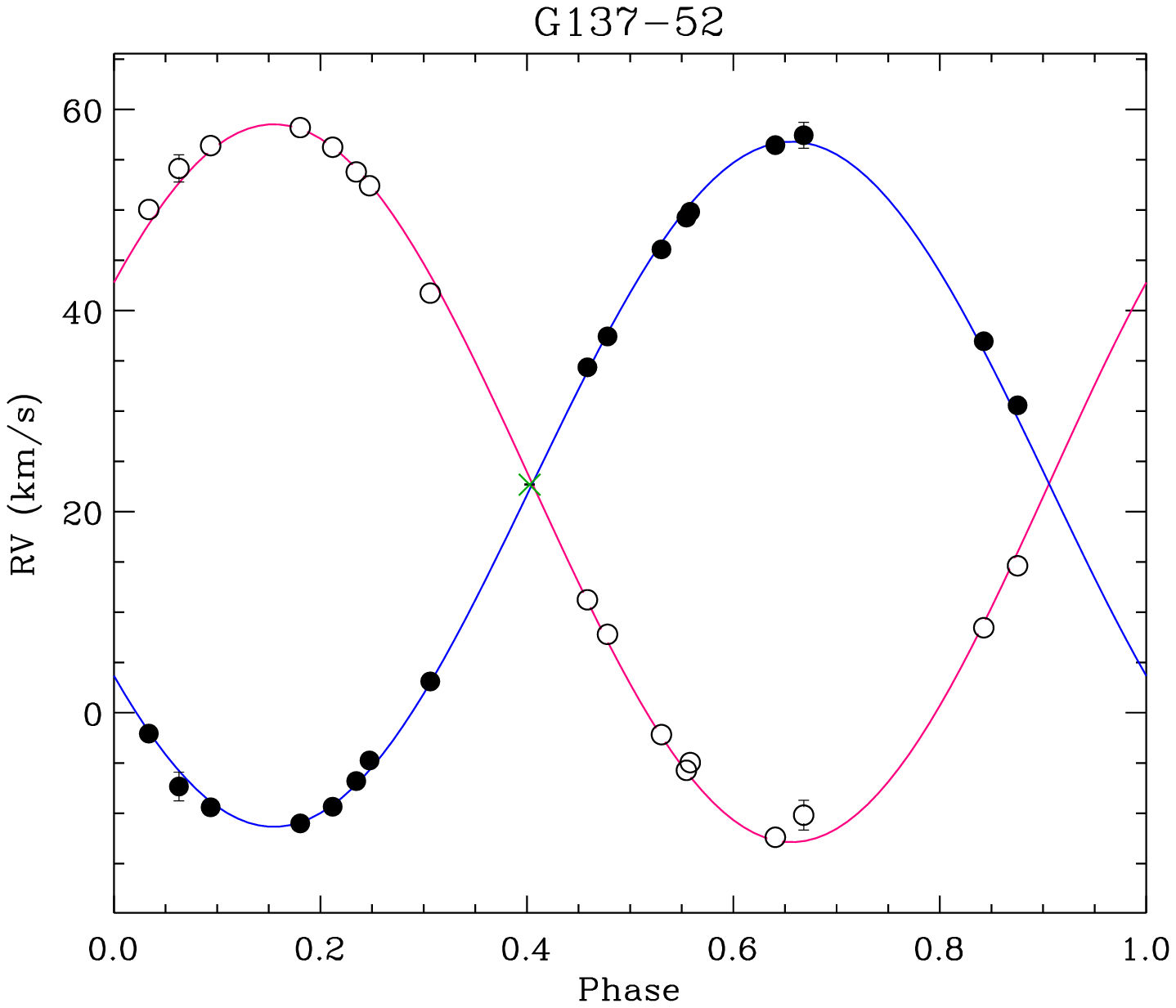}	 
  \includegraphics[width=5.5cm,height=5.5cm]{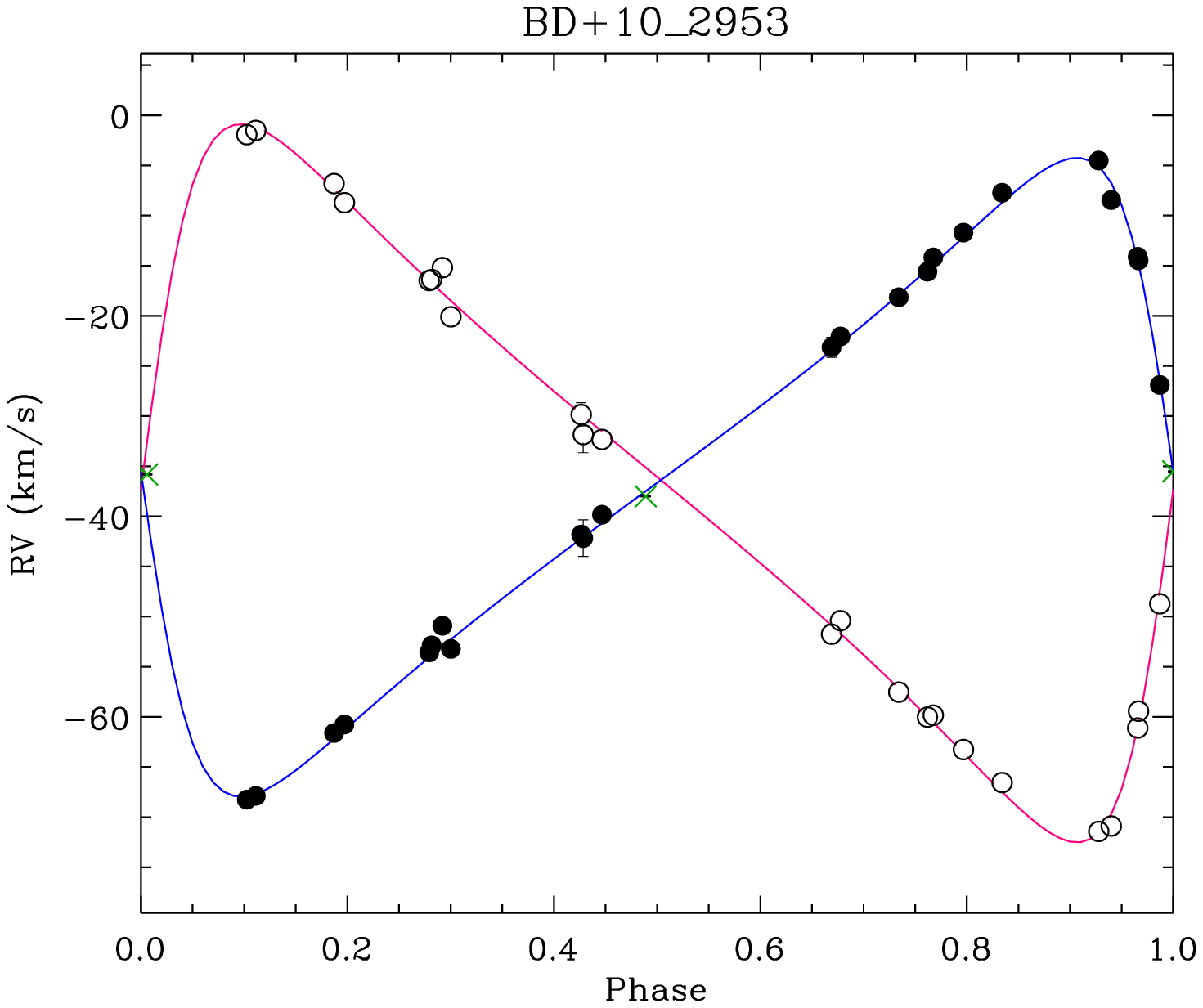}  
  \includegraphics[width=5.5cm,height=5.5cm]{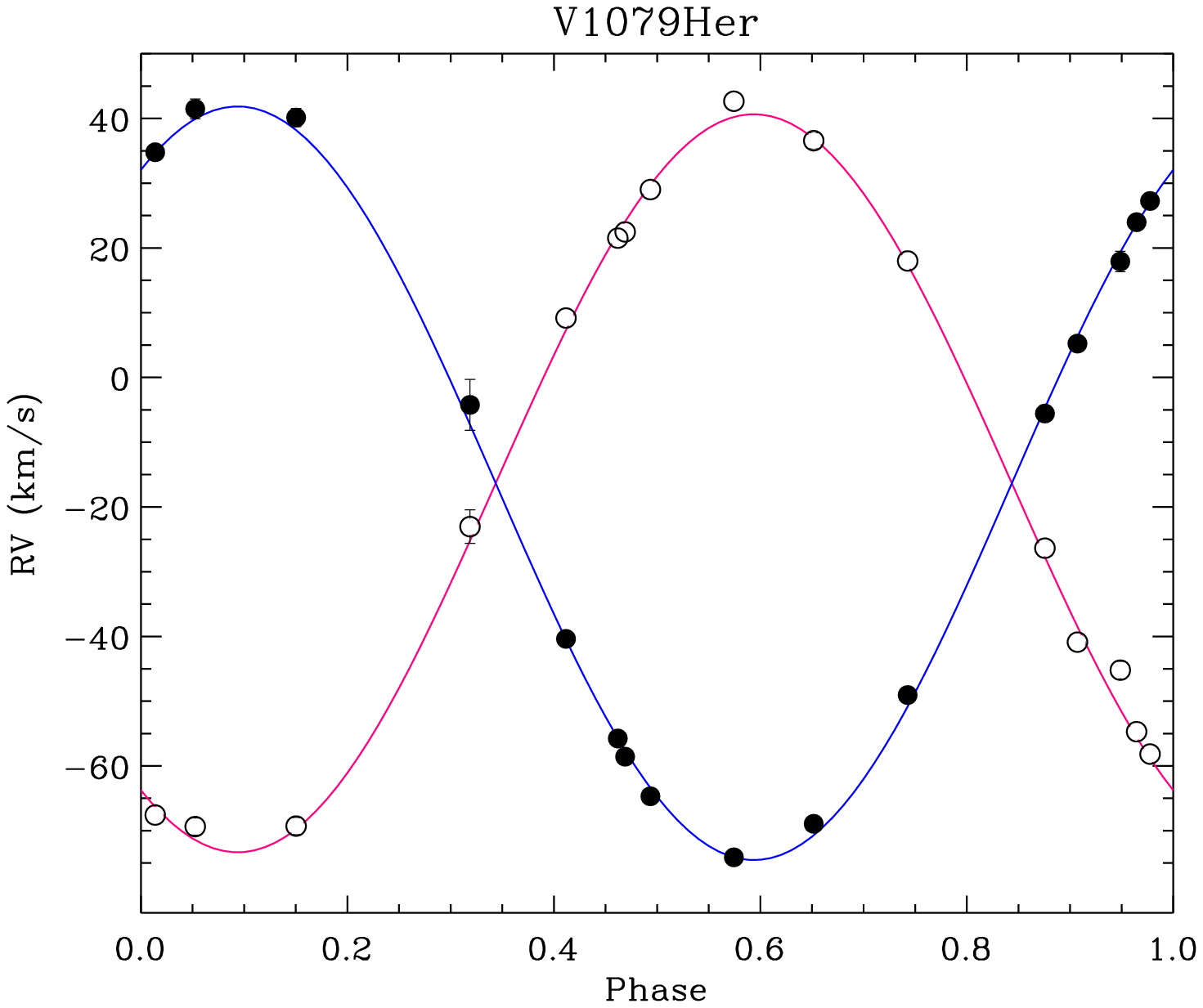}  
  \includegraphics[width=5.5cm,height=5.5cm]{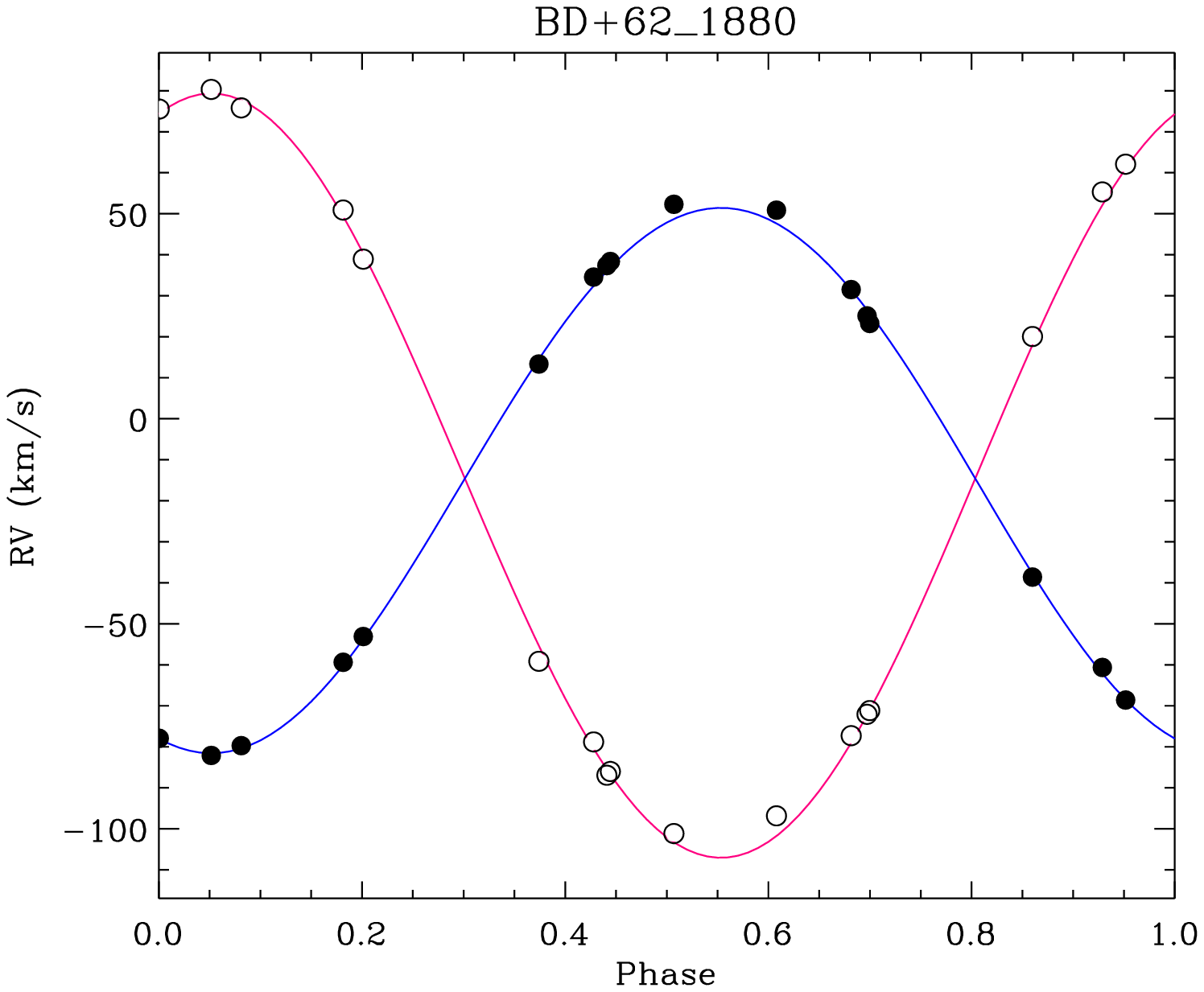}  
  \includegraphics[width=5.5cm,height=5.5cm]{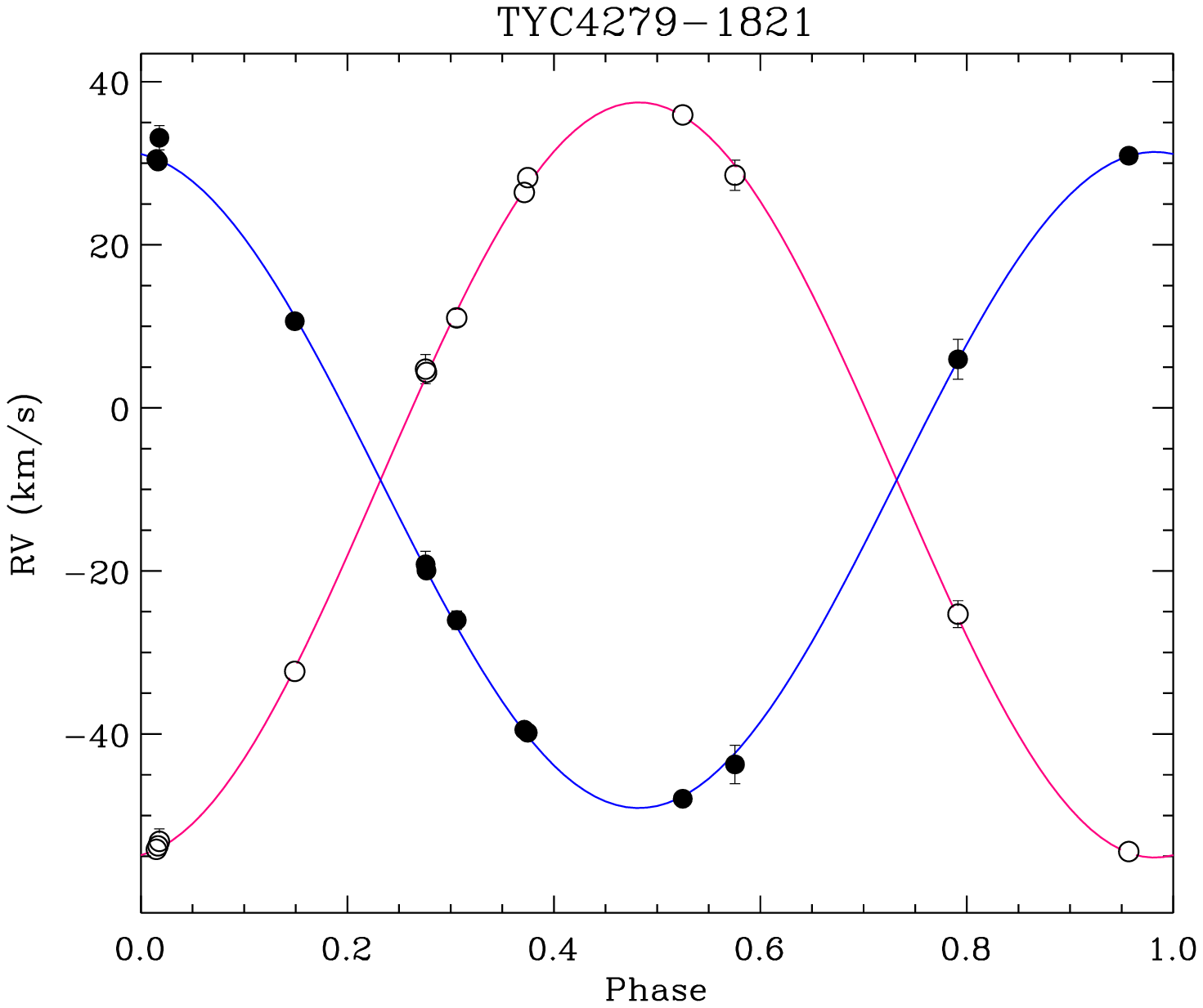}  
  \end{center}
\caption{Radial velocity curves of the six {\it RasTyc} binaries. 
Filled and open symbols for the primary (more luminous) and secondary components have been used, respectively. The green 
crosses relate to the RV measured near the conjunctions, when only one peak is visible in the cross-correlation function. 
The blue and red lines represent the orbital solutions for the primary (more luminous) and secondary components, respectively.  }
\label{fig:RV}
\end{figure*}

\begin{table*}	%[ht]
\caption[Orbital parameters of the {\it RasTyc} binaries.]%<-- this version will appear in List of Tables
{Orbital parameters of the {\it RasTyc} binaries.}%<-- this version will appear on page
\begin{tabular}{lcccccccc}
\hline
\noalign{\smallskip}
   Name   & HJD0$^a$                    &  $P_{\rm orb}$ & $e$   &  $\omega$    &  $\gamma$      &      $k$       &  $M\sin^3i$   & $M_{\rm P}$/$M_{\rm S}$ \\
          & {\scriptsize(2\,450\,000+)} & (days)	 &       & ($^{\circ}$) & (km\,s$^{-1}$) & (km\,s$^{-1}$) & (M$_{\odot}$) &              \\
          &                             &                &       &              &                &     [P/S]      &     [P/S]     &              \\
\noalign{\smallskip}
\hline
\noalign{\smallskip}
 TYC\,3386-868-1  & 4453.68(5)   & 13.8204(1)   & 0.015(5)  &  88(5)   &    23.8(2)  & 61.4(3)/66.1(2)    & 1.54(1)/1.43(1)     & 1.077(5)  \\
 G\,137-52        & 2436.02(4)   & 5.04823(1)   & 0.027(1)  & 124(3)   &    22.77(2) & 34.07(4)/35.70(7)  & 0.0910(4)/0.0868(3) & 1.048(2)  \\
 BD+10\,2953      & 4105.33(3)   & 33.5251(3)   & 0.510(1)  & 269.0(2) & $-$36.38(2) & 31.86(3)/35.82(5)  & 0.363(1)/0.323(1)   & 1.124(2)  \\
 V1079~Her        & 2421.0726(1) & 19.4085(1)   & 0.000(1)  & 326.3(1) & $-$16.34(6) & 58.2(1)/57.0(2)    & 1.524(9)/1.556(7)   & 0.980(3)  \\
 BD+62\,1880      & 7258.92(2)   & 3.79074(1)   & 0.008(3)  & 161(1)   & $-$14.57(5) & 66.49(8)/93.20(24) & 0.936(5)/0.667(3)   & 1.402(2)  \\
 TYC\,4279-1821-1 & 2214.4952(6) & 15.55528(1)  & 0.0000(2) &  6.46(2) & $-$8.85(6)  & 40.24(8)/46.31(11) & 0.560(3)/0.487(2)   & 1.151(3)  \\
\noalign{\smallskip}
\hline \\
\end{tabular}
\label{Tab:RV}
\begin{list}{}{}
\item[] The errors on the last significant digit are enclosed in parenthesis. P=Primary (more luminous), S=Secondary.
\item[$^{\rm a}$] Heliocentric Julian Date (HJD) of the periastron passage.
\end{list}
\end{table*}

\subsection{Radial velocity}	
\label{sec:RV}

The radial velocity (RV) was measured by cross-correlating targets and template spectra. The latter are ATLAS9  \citep{Kurucz1993}
synthetic spectra with a solar metallicity and \teff\ in the range 4000--6000\,K that are calculated with SYNTHE \citep{Kurucz1981} at the same resolution and sampling of the CAOS ones.  
For this purpose we used the IRAF task {\sc fxcor} \citep[][]{Tonry1979,Fitzpatrick1993}. 
We excluded broad spectral features, such as Balmer and Na{\sc i} D$_2$ lines, because they blur	
the cross-correlation function (CCF) and hamper the measure of the RVs of the two components.
The spectral ranges heavily affected by telluric absorption lines were discarded as well.	

For each spectrum, the centroids of the CCF peaks of the two components are obtained by fitting two Gaussians, which
allows us to disentangle the two peaks in cases of partial line blending.  At orbital phases very close to the conjunctions, where the spectral lines of the two components are fully superposed, only one CCF peak is visible and the single Gaussian fitting provides a ``blended'' RV value, which is close to the barycentric velocity of the system. 

The radial velocity measurements of these binaries are reported in Tables~\ref{Tab:RV_TYC3386}--\ref{Tab:RV_TYC4279} 
along with their errors ($\sigma_{RV}$). The latter were computed by {\sc fxcor} according to the fitted peak height
and the antisymmetric noise as described by \citet{Tonry1979}.
The observed RV curves  are displayed in Fig.~\ref{fig:RV}, where we used filled symbols for the primary (more luminous) components and open symbols for the secondary ones. 
With the exception of V1079\,Her, which is composed of very similar stars with a mass ratio close to 1, the primary components are also 
the more massive ones.
The RVs obtained in fully blended situations are marked with green crosses in Fig.~\ref{fig:RV}.	%and were not used for the RV solutions.
%We initially searched  for eccentric orbits  and, in any case, found very low eccentricity values (e.g., $e=0.01$
%for HD~183957, $e=0.03$ for BD+33\,4462). Thus, following the precepts of \citet{Lucy1971}, we adopted $e=0$.
We used periodogram analysis \citep{Scargle1982} and the CLEAN deconvolution algorithm \citep{Roberts1987}, which allowed us to reject aliases generated by the spectral window of the unevenly sampled data, 
to determine the orbital periods from the RV variations of the SB2's components. 
%The data folded with the period display a smooth variation with an asymmetrical shape typical for an eccentric RV orbital motion (see Fig. 4a). 
Then, we fitted the observed RV curve with the {\sf IDL}\footnote{IDL (Interactive Data Language) is a registered trademark of  Harris Corporation.} 
routine {\sc curvefit} \citep[e.g.,][]{Bevington}, adopting the function {\sc helio\_rv} for spectroscopic binaries with eccentric orbits, to determine 
the orbital parameters and their standard errors. 
The RV curve fitting also allowed us to improve the determination of the orbital period.
The orbital solutions are overplotted to the RV data in Fig.~\ref{fig:RV}. 

The orbital period ($P_{\rm orb}$), barycentric velocity ($\gamma$), eccentricity ($e$), longitude of periastron ($\omega$), RV semi-amplitudes ($k$), 
masses ($M\sin^3i$), and mass ratios ($M_{\rm P}$/$M_{\rm S}$) for each binary system are listed in Table~\ref{Tab:RV}, where P and S  refer to the primary (more 
luminous) and secondary components of the SB2 systems, respectively.

\begin{figure}
\begin{center}
\vspace{.5cm} 
\includegraphics[width=8.5cm,angle=0]{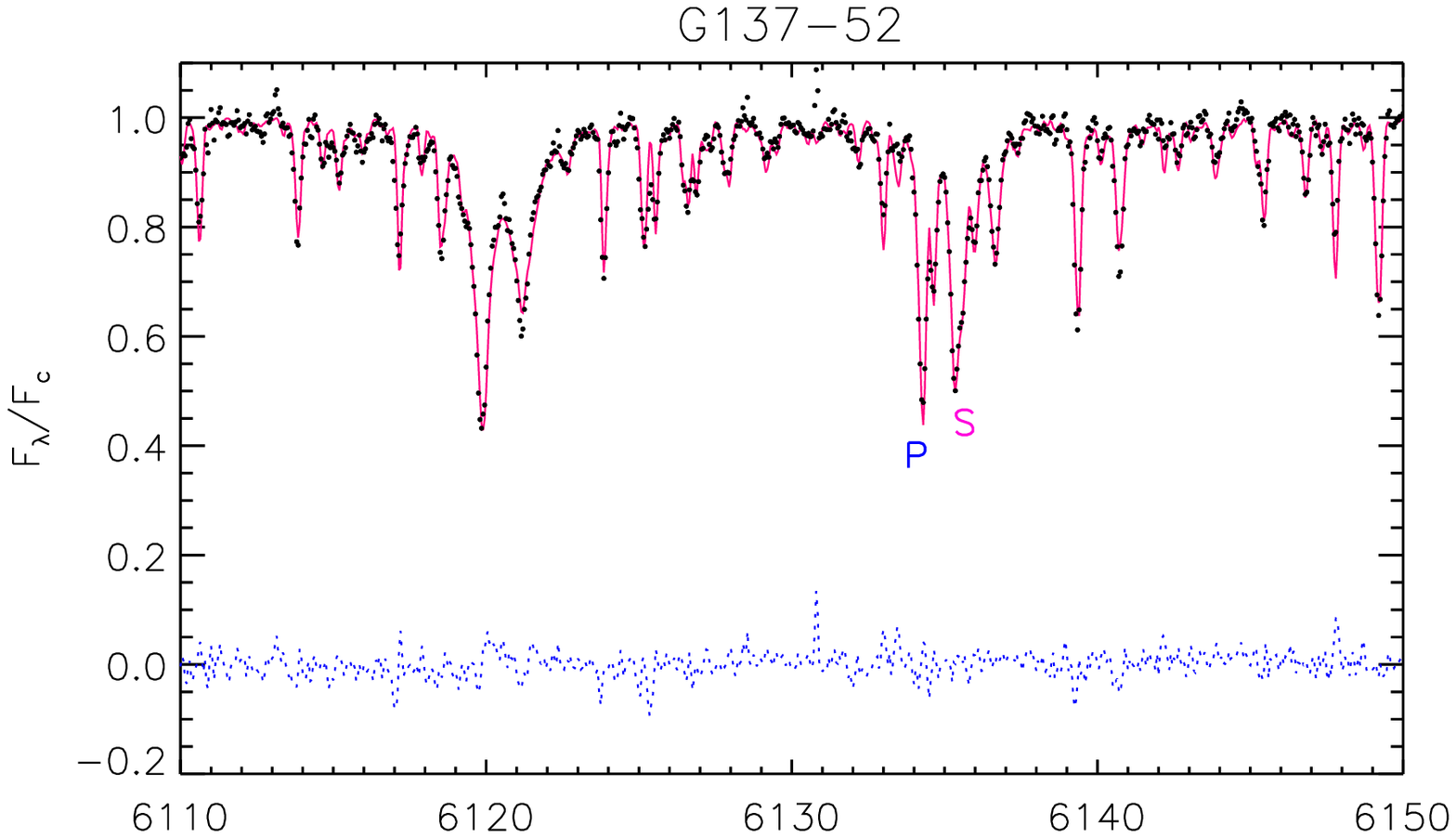}    
\vspace{-.5cm} 
\includegraphics[width=8.5cm,angle=0]{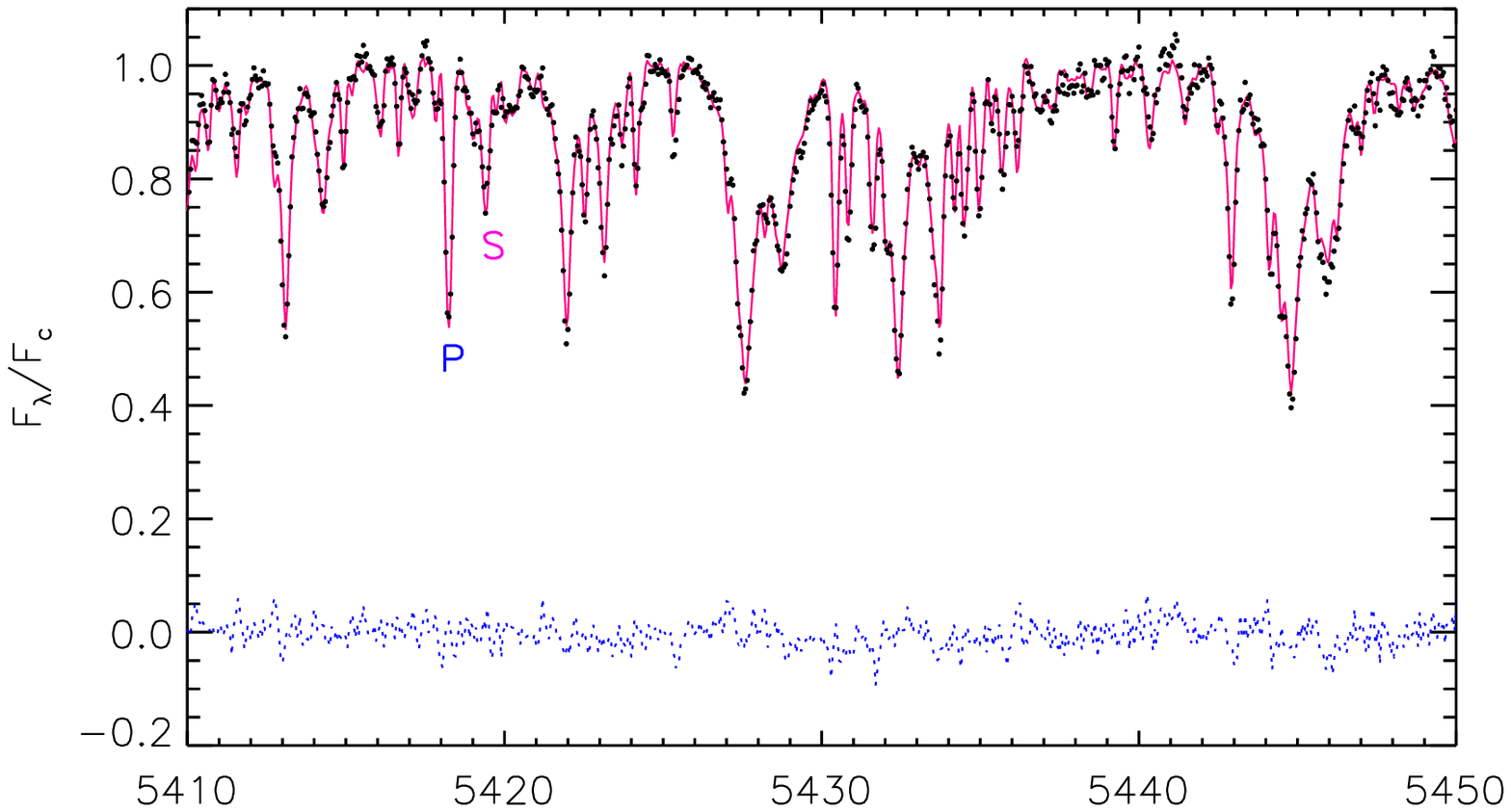}    
\vspace{-.5cm} 
\includegraphics[width=8.5cm,angle=0]{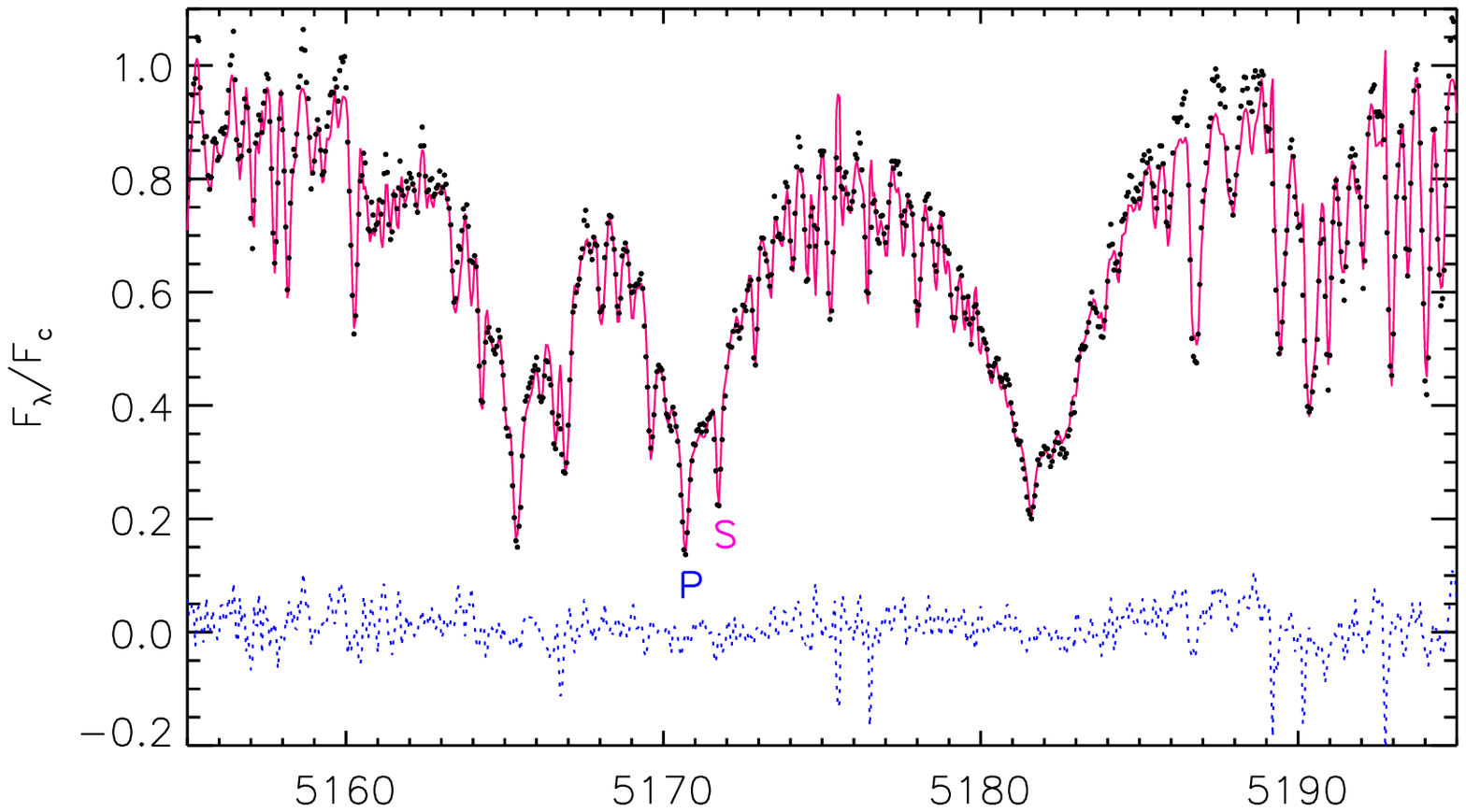}    
%\vspace{-.5cm} 
%\includegraphics[width=8.5cm,angle=0]{fig_G137_4900.eps}    
\caption{Example of spectral synthesis with \COMPO. In each panel the continuum-normalized spectrum of G\,137-52 is plotted with
black dots and it is overlaid to the synthetic spectrum (red lines) built up with the Doppler-shifted spectra of two standard 
stars mimicking the two components. The residuals are shown with blue dotted lines in the bottom of each panel. A few spectral 
lines of the primary and secondary component are also marked. } 
\label{Fig:COMPO2} 
\end{center}   
\end{figure}

\subsection{Atmospheric parameters}
\label{sec:APS}

For the determination of the atmospheric parameters (\teff\ and \logg) and to perform
an MK classification of the components of these binary systems we used \COMPO\ \citep{Frasca2006,Frasca2021}. This code was developed in 
{\sf IDL} environment and uses a grid of templates to reproduce the observed composite spectrum. As templates we adopted spectra of slowly
rotating, low-activity stars retrieved from the ELODIE Archive \citep{Moultaka2004}, whose atmospheric parameters are known from
the literature and listed in the PASTEL catalogue \citep{PASTEL}. The grid is composed of spectra of 90 stars both on the main sequence 
(FGKM types) and G-K giants.

For each binary system, we have chosen the spectrum with the best SNR and a large wavelength separation between the lines of the components, 
as derived from the CCF analysis. We have analyzed 39 \'echelle orders that cover the wavelength range 4290--6720~\AA. 
The projected rotation velocities, \vsini$_1$ and \vsini$_2$, were measured by \citet[][table A.2]{Frasca2018} from the 
FWHM of the peaks of the CCF and are kept fixed in the fit. 
The rotationally-broadened templates are shifted in wavelength according to the RV measured as described in Section~\ref{sec:RV} and summed, after weighting them according to their contribution to the local continuum. 
%The RV separation of the two templates that reproduce the two components in the composite spectrum is measured from the CCF peaks as described 
%in Section~\ref{sec:RV} and is also kept fixed in \COMPO. 
The flux ratio, i.e. the flux contribution of the primary component in units of the continuum, $w^{\rm P}$, is  an adjustable parameter. 
An example of the application of \COMPO\ is shown in Fig.~\ref{Fig:COMPO2} for three spectral segments of a spectrum of G\,137-52.

To evaluate the atmospheric parameters (APs), per each spectral segment we kept, among the 8100 possibilities, only the best 100 combinations 
(in terms of minimum $\chi^2$) 
of primary and secondary spectra. 
To calculate the average APs, the results for individual segments have been weighted with the $\chi^2$ of the fit and the total line absorption, 
$f_i=\int(F_{\lambda}/F_{\rm C}-1)d\lambda$, where $F_{\lambda}/F_{\rm C}$ is the continuum-normalized spectrum in the i-th
spectral segment. This parameter is an index of the amount of spectral information contained in each segment.
Usually, the blue spectral regions have more and deeper lines than the red ones (higher $f_i$), but their SNR and the $\chi^2$ of the fit are worse,
which reduces their weight in the mean.
 
\begin{figure}
\begin{center}
\hspace{-.3cm}
\includegraphics[width=9.cm]{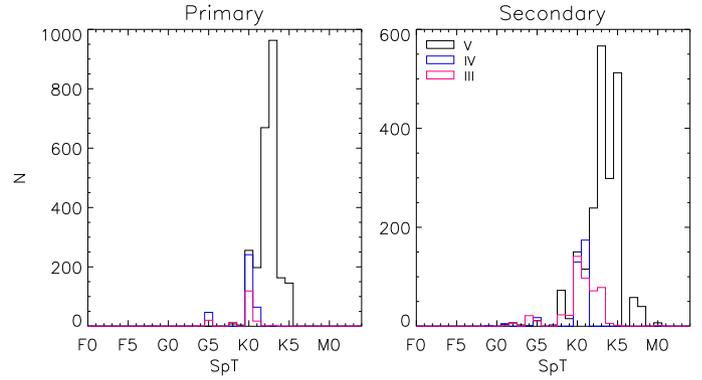}
\vspace{-0.2cm}
\caption{Distribution of spectral types for the components of G\,137-52.}
\label{fig:histo}
\end{center}
\end{figure}

\begin{figure}
\begin{center}
\hspace{-.3cm}
\includegraphics[width=9.cm]{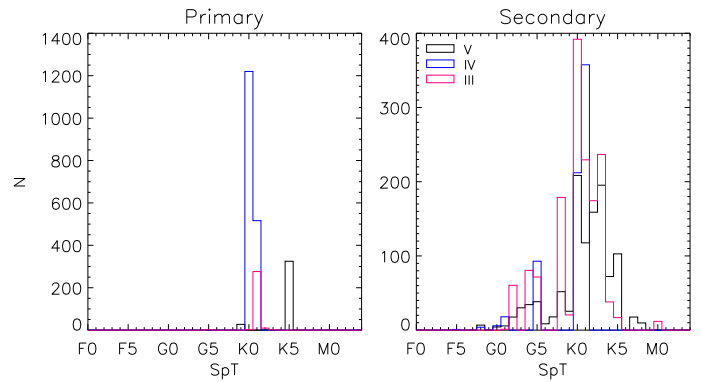}
\vspace{-0.2cm}
\caption{Distribution of spectral types for the components of V1079~Her.}
\label{fig:histo1}
\end{center}
\end{figure}

The APs are listed in Table~\ref{Tab:phys}. 
Averages values of $w^{\rm P}$ have been evaluated at three wavelengths (4600\,\AA, 5500\,\AA, and 6400\,\AA) by keeping
%using 
only the spectral segments around these wavelengths. 
As can be seen in Table~\ref{Tab:phys}, $w^{\rm P}$ changes appreciably with wavelength for systems with components 
of very different \teff, like TYC\,3386-868-1,  BD+10\,2953, TYC\,4279-1821-1, and BD+62\,1880. For the latter, the flux of the primary 
component is much larger than the secondary one at all wavelengths.
The spectral types (SpT) of the components are taken as the mode of the spectral-type distributions (Figs.~\ref{fig:histo} and \ref{fig:histo1} for 
two examples of binaries with MS and evolved components, respectively).

\begin{table*}	%[ht]
\caption[Physical parameters of the systems' components.]%<-- this version will appear in List of Tables
{Physical parameters of the systems' components derived in the present work. \vsini\ values are from \citet{Frasca2018}.}%<-- this version will appear on page
\begin{tabular}{lcccccccc}
\hline
\noalign{\smallskip}
Name  	  &  &  $v\sin i$  &  SpT  & $w_{4600}^{\rm P}$ & $w_{5500}^{\rm P}$  & $w_{6400}^{\rm P}$ &  $T_{\rm eff}$		  & \logg      \\ 
            	  &  & (km\,s$^{-1}$)&  	   &	                &		      & 		   &   (K)			  & (dex)    \\ 
            	  &  &  [P/S]	     &   [P/S]     &	                &		      & 		   &   [P/S]			  & [P/S]   	  \\ 
\noalign{\smallskip}
\hline
\noalign{\smallskip}
TYC\,3386-868-1  &  &   28.7/4.9    & K1III/G5IV  &  0.55\,$\pm$\,0.05 &  0.64\,$\pm$\,0.05  &  0.77\,$\pm$\,0.05 &   4870$\pm$140/5590$\pm$170  & 2.93$\pm$0.33/4.05$\pm$0.26 \\
 G\,137-52	  &  &   $<5/<5$     & K3V/K4V     &  0.65\,$\pm$\,0.03 &  0.64\,$\pm$\,0.02  &  0.62\,$\pm$\,0.03 &   4920$\pm$100/4850$\pm$100  & 4.57$\pm$0.14/4.50$\pm$0.28 \\ 
 BD+10\,2953	  &  &   10.5/5.0    & K0IV/G5IV   &  0.54\,$\pm$\,0.03 &  0.60\,$\pm$\,0.05  &  0.65\,$\pm$\,0.15 &   4980$\pm$130/5430$\pm$200  & 3.34$\pm$0.66/3.87$\pm$0.38  \\
 V1079~Her	  &  &   18.2/19.6 & K0IV/K1III-IV &  0.64\,$\pm$\,0.03 &  0.63\,$\pm$\,0.02  &  0.59\,$\pm$\,0.04 &   4780$\pm$140/4930$\pm$160  & 3.04$\pm$0.29/3.50$\pm$0.82  \\ 
 BD+62\,1880	  &  &   11.4/19.0   & F9V/K0V     &  0.90\,$\pm$\,0.07 &  0.85\,$\pm$\,0.05  &  0.80\,$\pm$\,0.07 &   6100$\pm$~70/5000$\pm$290  & 4.16$\pm$0.11/4.22$\pm$0.47  \\
 TYC\,4279-1821-1 &  &   10.6/8.2    & K0IV/G5IV   &  0.43\,$\pm$\,0.05 &  0.57\,$\pm$\,0.05  &  0.67\,$\pm$\,0.06 &   4780$\pm$150/5500$\pm$230  & 3.21$\pm$0.52/4.14$\pm$0.29  \\ 
\hline \\
\end{tabular}
%\begin{list}{}{}								       
%\item[$^a$] From \citet{Frasca2018}.
%\end{list}
\label{Tab:phys}
\end{table*}

\subsection{Hertzsprung-Russel diagram}
\label{sec:HR}

In order to evaluate the consistency of the dynamical masses, previously determined, and the evolutionary ones, inferred from theoretical models, we constructed the Hertzsprung-Russel (HR) diagram.

We adopted the temperatures derived with \COMPO\ and listed in Table~\ref{Tab:phys}.
The luminosities of the components of the SB2 systems, $L_{\rm P,S}$, were derived from the combined $V$ magnitude listed in Table~\ref{Tab:X} and the luminosity ratio at 5500\,\AA\ (which is related to $w^{\rm P}_{5500}$ in Table~\ref{Tab:phys}). The values of $V$ have been corrected for the extinction, $A_{\rm V}$, and have been used to calculate the absolute magnitudes in the $V$ band, $M_V$, with the \gaia\ distances ($d$).
The latter have been estimated by direct inversion of the parallax, after having been corrected according to the recommendations outlined by \citet{Lindegren2021}. Once distances and positions in the sky were known, the $A_{\rm V}$ for each system was determined from the maps of extinction obtained by \citet{Lallement2019}.
For these bright and relatively nearby sources, the values of extinction  are rather low ($A_V=0.01-0.31$\,mag, Table~\ref{Tab:HR}).

The bolometric correction of \citet{PecautMamajek2013} was applied to get the bolometric magnitudes from the values of $M_V$. 
Finally, the bolometric magnitude of the Sun, $M_{\rm bol}^{\odot}=4.64$\,mag \citep{Cox2000}, was used to express the stellar luminosity in solar units.
The error of luminosity includes the parallax error, the error on $w^{\rm P}_{5500}$, and the uncertainty on the $V_0$ magnitude. The latter includes the
error on the extinction, which has been estimated as 0.2 mag.  

\begin{table*}	%[ht]
\caption[Physical parameters inferred from the HR diagram.]%<-- this version will appear in List of Tables
{Physical parameters related to or inferred from the HR diagram.}%<-- this version will appear on page
\begin{tabular}{lrcrrccc}
\hline
\noalign{\smallskip}
          Name  &  $d$~~~ &	$A_V$ &	 $L_{\rm P}$~~~~~~ & 
        $L_{\rm S}$~~~~~~  & $M_{\rm P}$ & $M_{\rm S}$ & $i^a$ \\
                &   (pc)~  & (mag) &    (\Lsun)~~~~ &  (\Lsun)~~~~ & (\Msun)  & (\Msun) &  ($\degr$) \\
\noalign{\smallskip}
\hline 
\noalign{\smallskip}
TYC\,3386-868-1  &   293.6 &  0.257 &	13.482$\pm$3.252  &   6.069$\pm$1.837 & 1.40$^{+0.20}_{-0.30}$  & 1.38$^{+0.10}_{-0.15}$ & 90 \\
        G\,137-52  &    55.5 &  0.006 & 0.267$\pm$0.058  &   0.185$\pm$0.047 & 0.82$^{+0.05}_{-0.10}$  & 0.77$^{+0.05}_{-0.10}$ & 30 \\
     BD+10\,2953  &   224.2 &  0.044 & 6.511$\pm$1.476  &   3.508$\pm$1.012 & 1.35$^{+0.15}_{-0.25}$  & 1.18$^{+0.12}_{-0.13}$ & 40 \\
      V1079~Her  &   404.7 &  0.169 & 21.114$\pm$4.257  &  11.391$\pm$2.937 & 1.42$^{+0.30}_{-0.30}$  & 1.45$^{+0.20}_{-0.25}$ & 90  \\
     BD+62\,1880  &   125.8 &  0.014 & 1.320$\pm$0.241  &   0.290$\pm$0.133 & 1.12$^{+0.05}_{-0.05}$  & 0.83$^{+0.08}_{-0.15}$ & 70 \\
 TYC\,4279-1821-1  &   314.8 &  0.313 &	10.535$\pm$2.212  &   4.007$\pm$1.152 & 1.10$^{+0.30}_{-0.25}$  & 1.20$^{+0.15}_{-0.10}$ & 50  \\
\hline \\
\end{tabular}
\begin{list}{}{}								       
%\item[$^a$] From \citet{Frasca2018}.
\item[$^a$] Derived from the comparison between the evolutionary masses, $M_{\rm P,S}$, inferred from the position on the HR diagram  and the dynamical masses listed in Table~\ref{Tab:RV}.
\end{list}
\label{Tab:HR}
\end{table*}

\begin{figure}
\hspace{-0.4cm} 
%\begin{center}
\includegraphics[width=8.7cm,angle=0]{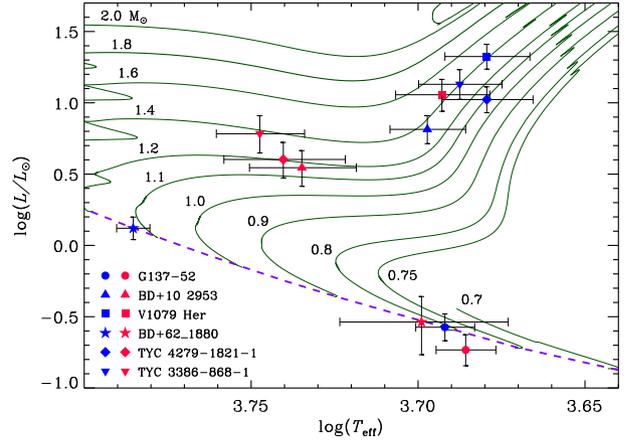}    
\caption{HR diagram. Blue and red symbols are used for the primary and secondary components, respectively. Evolutionary tracks  with solar metallicity ($Z$=0.017) from \citet{Bressan2012} are overlaid by continuous lines. The Zero Age Main Sequence (ZAMS) at $\tau$=200\,Myr is shown by a purple dashed line.}
%The isochrone at $age = 200$\,Myr (ZAMS) is shown by a purple dashed line.
\label{fig:HR}
%\end{center}   
\end{figure}

The HR diagram for the components of the six {\it RasTyc} binaries, represented with different symbols, is shown 
in Fig.~\ref{fig:HR} where the PARSEC evolutionary tracks \citep{Bressan2012} for a solar metallicity (Z\,=\,0.017)
are displayed with continuous lines. 
This diagram confirms the MS stage for both components of G\,137-52 and  BD+62\,1880, in agreement with their MK spectral classification.
The secondary components of TYC\,3386-868-1, BD+10\,2953, and TYC\,4279-1821-1, which we classified all as G5IV, are correctly
located in the region of the HR occupied by subgiant stars, while the remaining stars lie in the red-giant branch. 
Their masses are in the range 1--2 \Msun.

Comparing the masses of the two components of each system, which can be inferred for their position on the HR diagram (quoted in Table~\ref{Tab:HR}), with the dynamical masses $M_{\rm P,S}\sin^3i$ (Table~\ref{Tab:RV}), we can derive the inclination, $i$, of these systems.
The values of $i$ derived from the two components of each system agree very well, within 3--5$\degr$, with each other. The average value is reported in Table~\ref{Tab:HR}. 
There are only two cases where the masses of the components derived from the HR diagram are smaller than 
(but nearly equal to) the dynamical masses $M_{\rm P,S}\sin^3i$, implying $\sin i>1$, namely TYC\,3386-868-1 and V1079~Her. In these cases we have quoted an inclination of 90$\degr$ 
in Table~\ref{Tab:HR}. However, the mass uncertainty of 0.15--0.30\,\Msun\ for these stars (Table\,\ref{Tab:HR}) allows for lower values of inclination down to $i\sim75\degr$, for which these systems with large separations should not display eclipses, in agreement with the behaviour of their light curves (see Section~\ref{Sec:Notes}).
Moreover, if we consider larger values of extinction, like those of $A_V$=0.47\,mag and 0.26\,mag reported by \citet{Gontcharov2018} for TYC\,3386-868-1 and V1079~Her, respectively, the inconsistency between dynamical and evolutionary masses disappears.

\section{Chromospheric emission and lithium content}
\label{Sec:chrom_lithium}

\begin{figure*}
\begin{center}
\hspace{-.5cm}
\includegraphics[width=9.0cm]{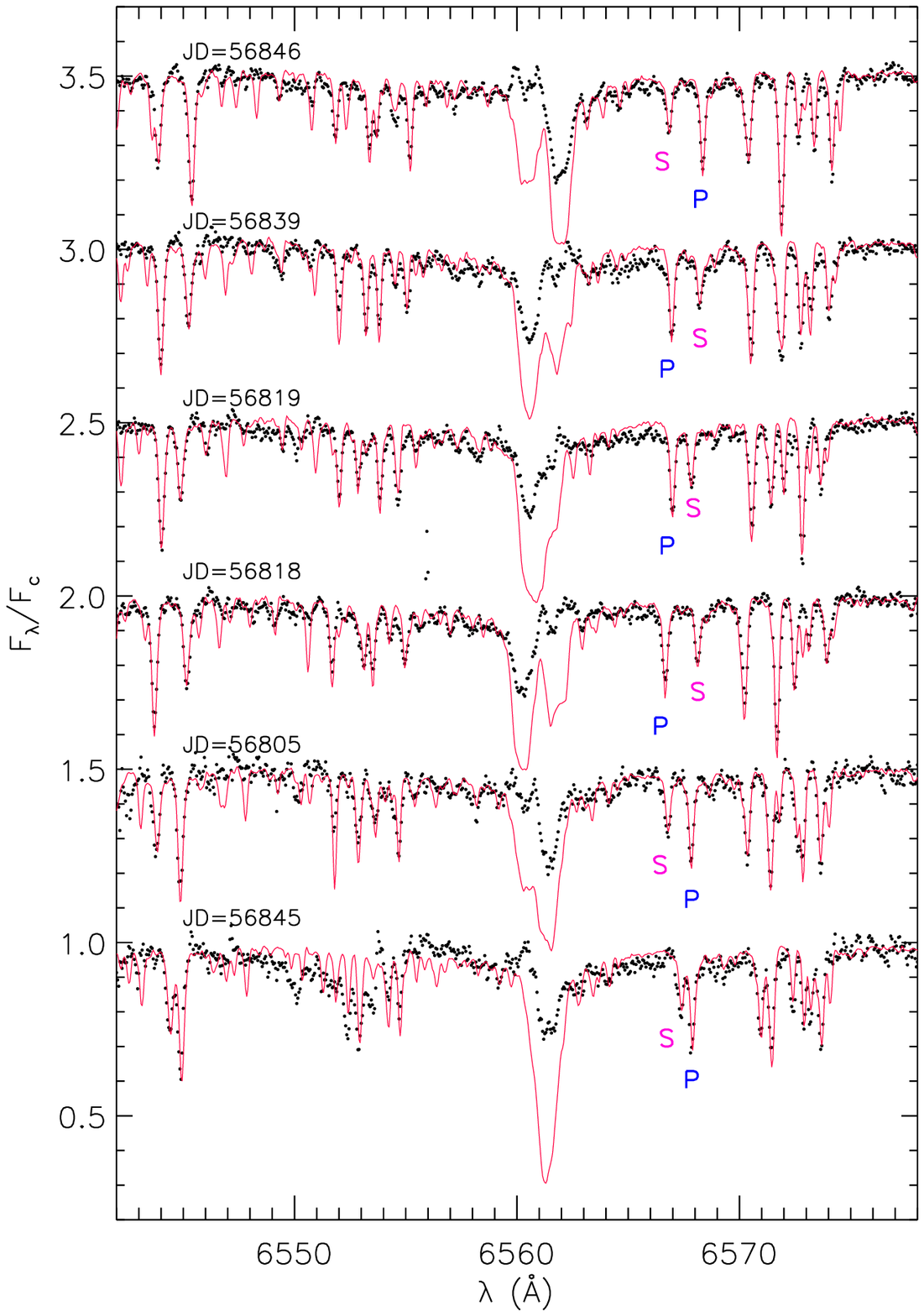}	%fig_subtract_ew_compo.ps}
\includegraphics[width=9.0cm]{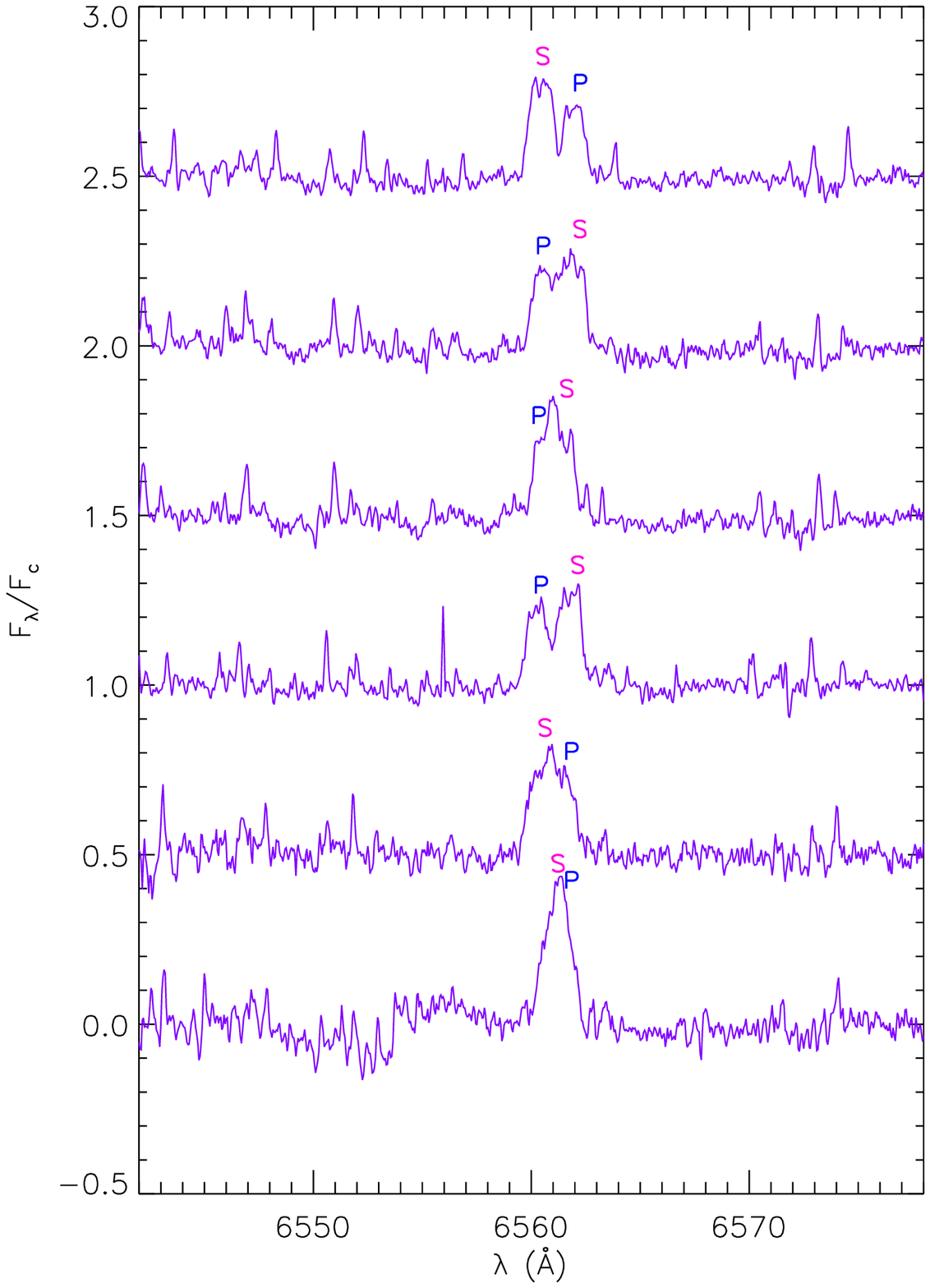}	%fig_subtract_ew_compo.ps}
\vspace{0cm}
\caption{Sample of CAOS spectra of G\,137-52 in the H$\alpha$ region taken at different orbital phases. 
{\it Left panel)} Synthetic composite spectra (red lines) are overlaid with the observed spectra (black dots) taken at the Julian date marked on top on each spectrum. The position of the Fe{\sc i}\,$\lambda$6569\,\AA\ line of the 
primary (P) and secondary (S) component is also marked in each spectrum. 
The chromospheric emission which fills in the H$\alpha$ core of both components is clearly visible in the subtracted 
spectra ({\it right panel}) where the H$\alpha$ wavelength of the primary and secondary component is also marked.}
\label{fig:subtraction_G137}
\end{center}
\end{figure*}

\begin{figure*}
\begin{center}
\hspace{-.5cm}
\includegraphics[width=5.7cm]{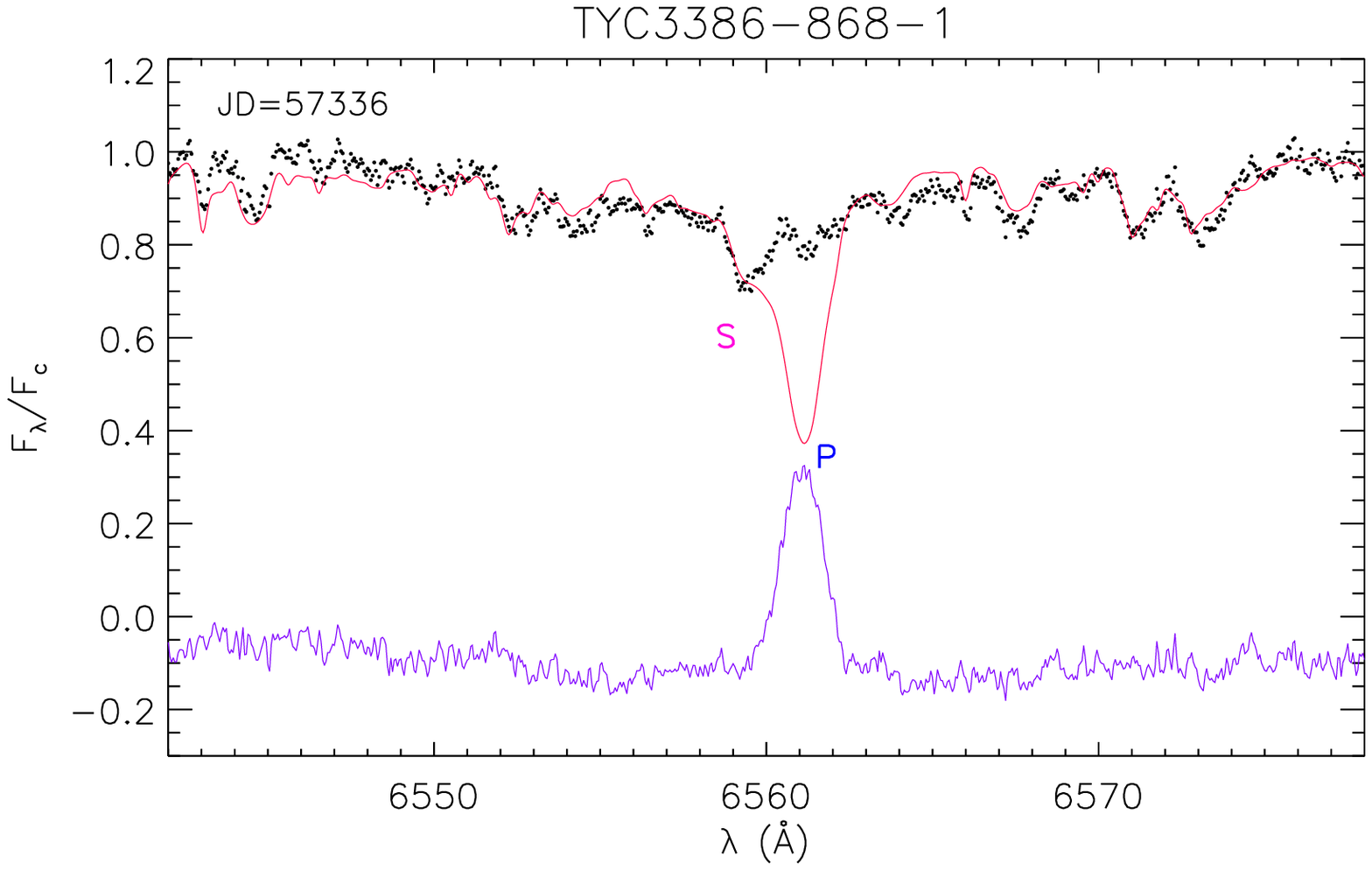}	
\includegraphics[width=5.7cm]{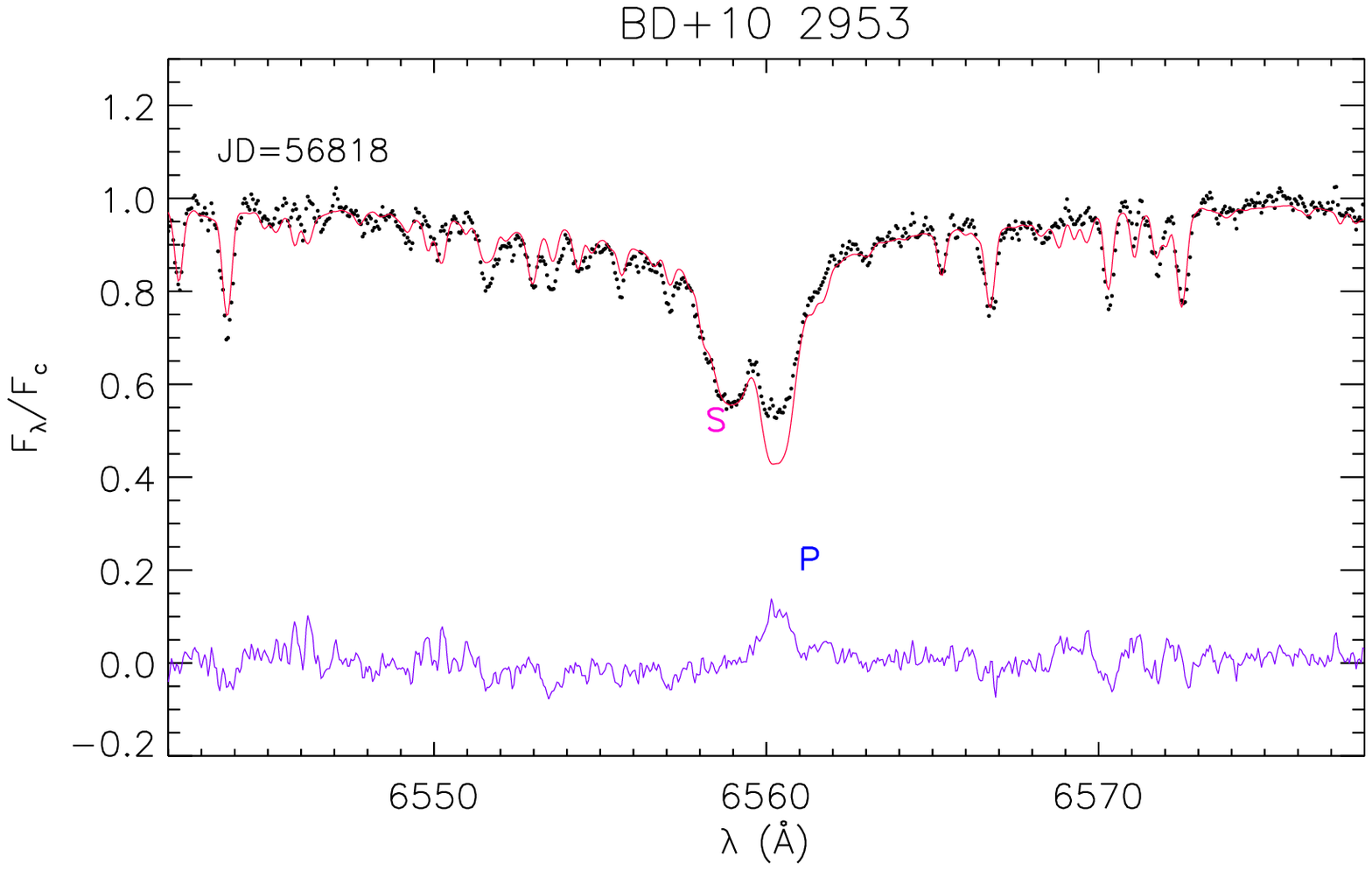}     
\includegraphics[width=5.7cm]{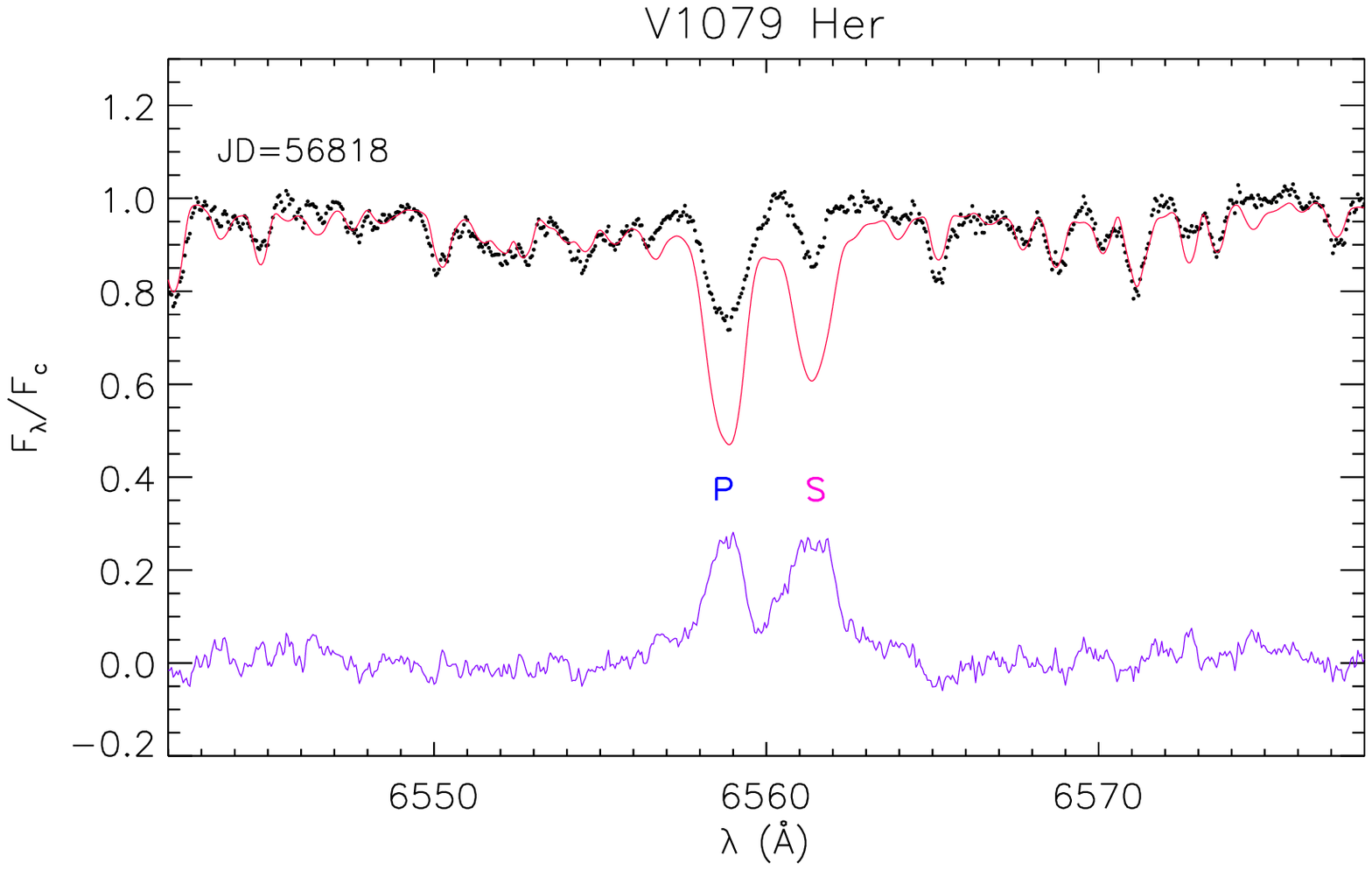}     
\includegraphics[width=5.7cm]{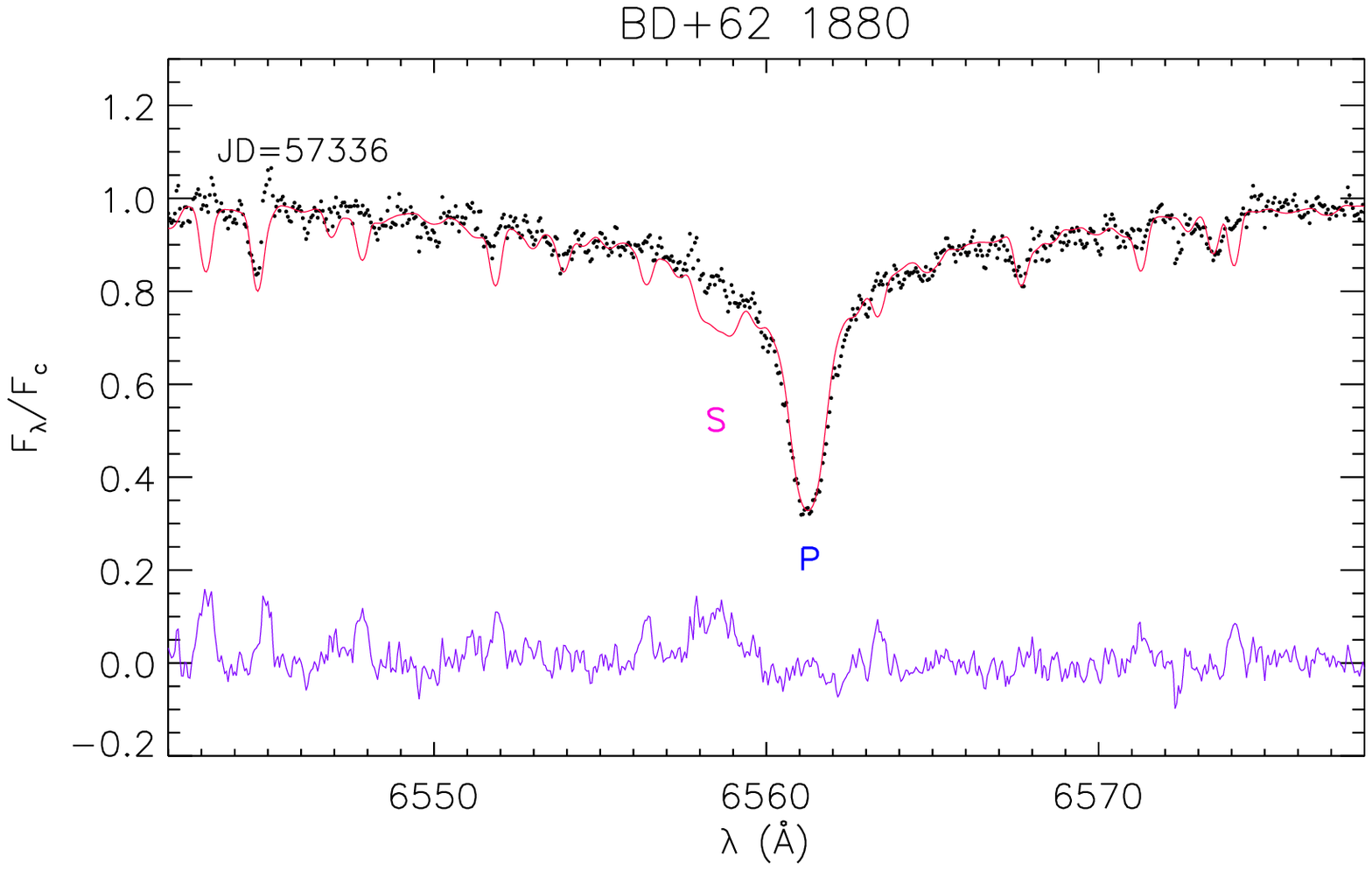}     
\includegraphics[width=5.7cm]{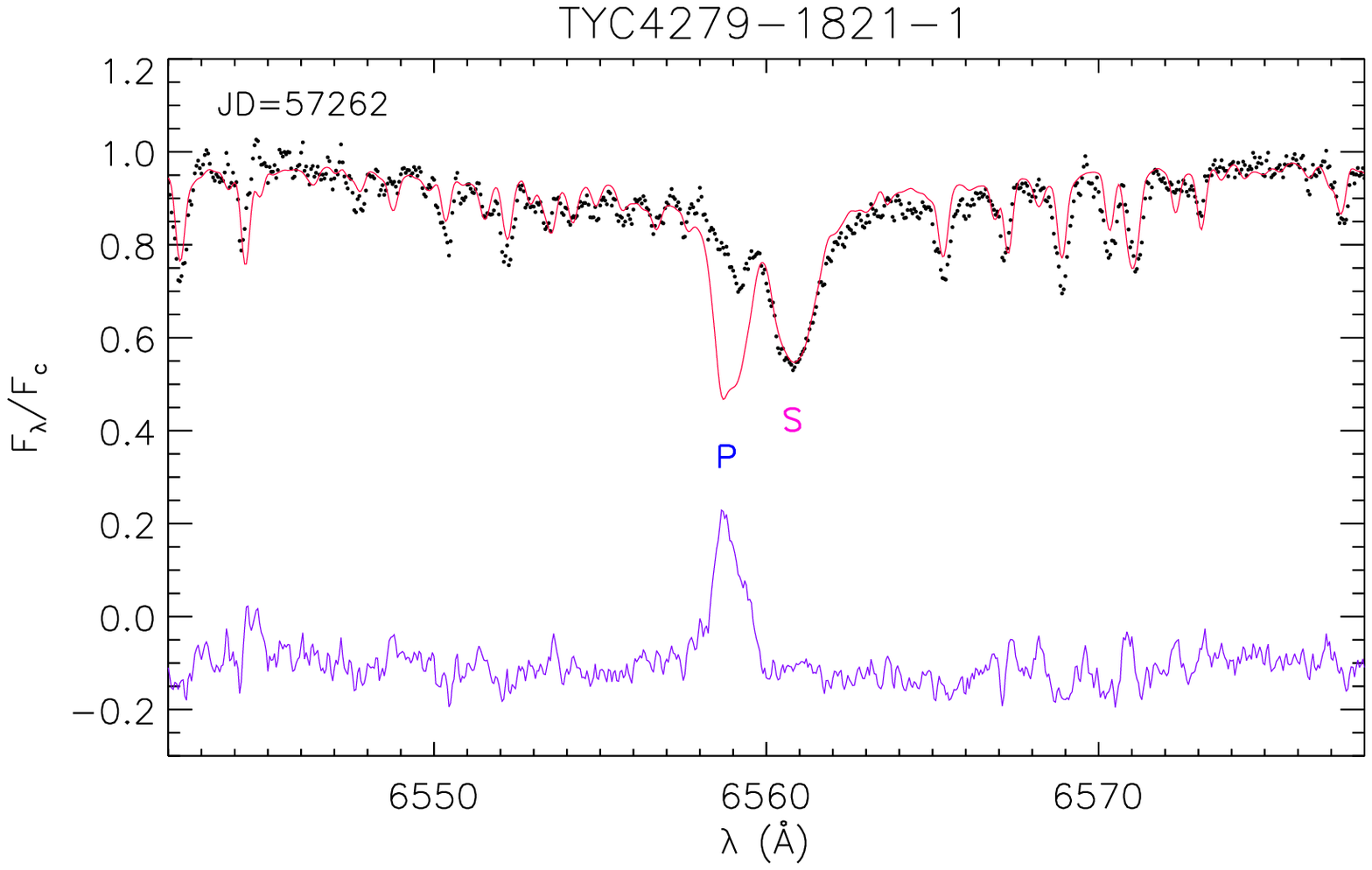}     
\vspace{0cm}
\caption{CAOS spectra of the investigated systems in the H$\alpha$ region. 
In each box, the synthetic composite spectra (red lines) are overlaid with the observed spectra (black dots). 
The position of the H$\alpha$ line of the primary (P) and secondary (S) components are also marked. 
The chromospheric emission which fills in the H$\alpha$ core of one or both components is clearly visible 
in the subtracted spectra (purple line in the bottom of each panel).}
\label{fig:subtraction_halpha}
\end{center}
\end{figure*}

\begin{figure}
\begin{center}
\hspace{-.5cm}
\includegraphics[width=9.0cm]{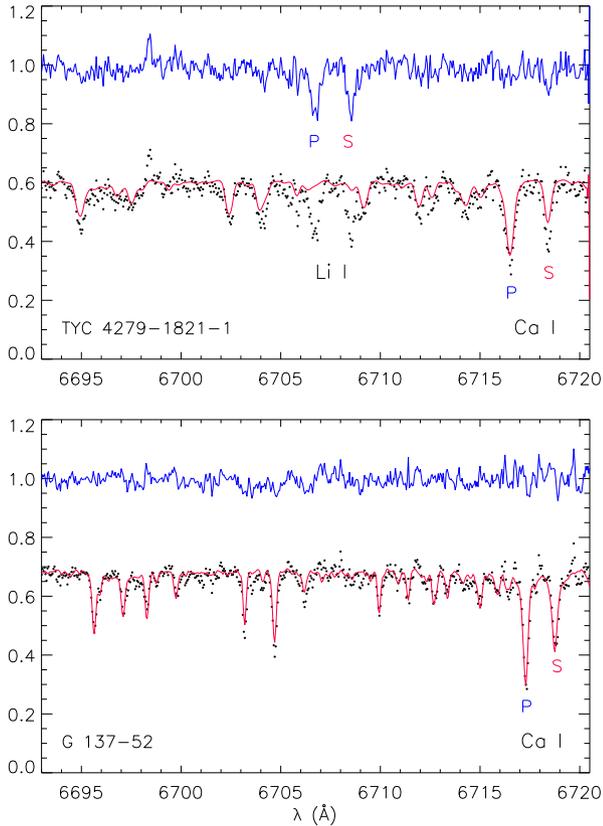}
\vspace{-2cm}
\caption{{\it Upper panel)} Subtraction of the synthetic composite spectrum (red line) from one observed spectrum of TYC 4279-1821-1 (black dots), which emphasizes the {Li}{\sc i} $\lambda$6708\,\AA\ absorption lines of the two components, marked near the residual spectrum (blue line) with `P' and `S' for the primary and secondary component, respectively. {\it Lower panel)} Observed (black dots) and synthetic composite spectrum (red line) of G\,137-52. The position of the {Ca}{\sc i} $\lambda$6717\,\AA\ lines of the primary and secondary component is marked. The {Li}{\sc i} lines of the two components are not detectable either in the observed or in the residual spectrum.}
\label{fig:subtraction_lithium}
\end{center}
\vspace{-.5cm}
\end{figure}

The level of chromospheric activity can be evaluated from the emission in the core of the H$\alpha$ line. 
The detection of chromospheric emission in the H$\alpha$ line core is not a trivial task for SB2 systems which display spectral lines of both components,  
with a different rotational broadening that are Doppler shifted at different wavelengths according to the orbital phase.  
To this end, we subtracted the composite spectra produced by \COMPO\ with non-active templates from the observed spectra of the targets, 
to remove the underlying photospheric lines so as to leave as residual the chromospheric emission that fills the H$\alpha$ cores of one or both components.
%The excess H$\alpha$ equivalent width, $EW_{\rm H\alpha}^{em}$, has been obtained by integrating the residual H$\alpha$ emission profile.
The same composite templates were also subtracted to the observed spectra to measure the equivalent 
width of the {Li}{\sc i}\,$\lambda$6708\,\AA\ absorption line ($EW_{\rm Li}$), removing the blends with nearby lines.
With the exception of TYC 4279-1821-1, which displays {Li}{\sc i} absorption from both components, for the remaining systems no lithium line could be clearly detected (see Fig.\,\ref{fig:subtraction_lithium}).  
Based on the noise in the residual spectrum, we estimate upper limits of about 30\,m\AA\ for $EW_{\rm Li}$ in the latter cases. In the best residual spectra of TYC 4279-1821-1 we have measured $EW_{\rm Li}=95\pm30$\,m\AA\ and $92\pm30$\,m\AA\ for the primary and secondary component, respectively.
These values must be divided by the contribution to the red continuum of the respective component,  $w_{6400}^{\rm P}=0.67$ and $w_{6400}^{\rm S}=0.33$, getting the correct values $EW_{\rm Li}^{\rm P}=140$\,m\AA\ and $EW_{\rm Li}^{\rm S}=280$\,m\AA.
We calculated the lithium abundance, $A$(Li), from our values of \teff, \logg, and $EW_{\rm Li}$ by interpolating the curves of growth of \citet{Lind2009}, which span the \teff\ range 4000--8000\,K and log from 1.0 to 5.0 and include non-local thermal equilibrium corrections. For the stars without Li{\sc i} detection, we have always found an abundance $A$(Li)$<2$, while we found $A$(Li)\,=2.2 and 3.1 for the primary and secondary component of TYC 4279-1821-1, respectively.
These abundances are larger than the typical values measured in single giants, but are in the range of those displayed by the so-called lithium-rich giants \citep[$A$(Li)$\geq$1.4, e.g.,][and reference therein]{Smiljanic2018,Martell21} and by some active binaries \citep[e.g.,][]{Pallavicini1992,Randich1993}. In particular, \citet{Randich1993,Randich1994} found that the evolved components of spectroscopic binaries present an excess Li abundance with respect to single stars of the same spectral type. However, they found high values of $A$(Li) only for a fraction of systems and moderate abundances ($A$(Li)$\leq$1.5) for most systems, with no obvious dependence on activity parameters such as rotation and chromospheric emission. Therefore, they suggest that activity {\it per se} is likely not the cause of the enhanced Li abundance.
%, lowering the importance of mechanisms such as the production of Li in large cool starspots or the production of Li in stellar flares by spallation reactions originally proposed by \citet{Pallavicini1992}.
This is also displayed by the four binaries of the present paper, whose evolved components have a comparably high level of chromospheric activity, but a high Li abundance has been measured only for TYC 4279-1821-1.
Anyway, the issue of atmospheric lithium overabundance in binaries is very complex also due to different effects related to evolution, activity, and binarity, which are simultaneously at work.
%For instance \citet{Barrado1997} found a Li overabundance for the MS components of chromospherically active binaries. They suggest that the angular momentum transfer from the orbit to the stellar rotation driven by tidal effects reduces the differential rotation and the related turbulent internal mixing with a consequent reduction of the rate of Li depletion. This is supported by the measure of lithium abundance in binaries belonging to open clusters, which display higher $A$(Li)n compared to the single stars of the same cluster 
For instance, \citet{Barrado1998} show that a significant part of the evolved components of the binaries studied by them have lithium excesses, independently of their mass and evolutionary stage.
They find instead the Li overabundance to be closely related to the stellar rotation, and interpret it as a consequence of the transfer of angular momentum from the orbit to the rotation as the stars evolve in and off the Main Sequence.
They suggest that the angular momentum transfer reduces the differential rotation and the related turbulent internal mixing with a consequent reduction of the rate of Li depletion. This is supported by the measure of lithium abundance in binaries belonging to the Hyades and M\,67 open clusters, which display higher $A$(Li) compared to the single stars of the same cluster. 
A relation of $A$(Li) with the stellar rotation for the components of active binaries, although with a large scatter, has been also found by \citet{Strassmeier2012}.  

Deriving accurate ages of pre-main sequence (PMS) and MS stars from the atmospheric lithium content is not a straightforward task \citep[e.g.,][]{Leone2007,Franciosini2022}. This task is even more complicated in the case of binary systems \citep[e.g.,][]{Giarrusso2016,Frasca2019}. A broad age classification for single stars and the components of SB2s can be instead carried out with the help of the upper envelopes of the lithium abundance distributions for members of open clusters \citep[e.g.,][and references therein]{Gutierrez2020}. 
We note that the upper limits of both $EW_{\rm Li}$ and $A$(Li) indicate that the MS components of the systems studied in the present work are older than about 1\,Gyr \citep[e.g.,][]{Randich2009,Jeffries2014,Gutierrez2020}, suggesting that the high level of chromospheric and coronal activity observed in these objects is not an age effect, but it is rather the result of the spin-orbit synchronization.

The spectral subtraction in the H$\alpha$ region reveals that both components of G\,137-52 are chromospherically active, with the less-massive 
cooler one displaying the stronger emission, which is just above the local continuum in most spectra (Fig.~\ref{fig:subtraction_G137}).
The same behaviour is displayed by V1079~Her which hosts two very active K-type giants (Fig.~\ref{fig:subtraction_halpha}).
For TYC\,3386-868-1, BD+10\,2953,and TYC\,4279-1821-1, which have the typical composition (K0-1\,IV-III/G5IV) of RS~CVn systems, the emission is clearly 
detected only for the cooler (brighter and more massive) components (Fig.~\ref{fig:subtraction_halpha}). 
The same holds true for BD+62\,1880, a system composed of main-sequence stars, where H$\alpha$ emission is only seen at the wavelength of the cooler K0V component, 
while the F9V primary does not show any filling in its H$\alpha$ core.

\section{Notes on individual objects}
\label{Sec:Notes}

\subsection{TYC\,3386-868-1}
This object is classified as a variable source in the All Sky Automated Survey for SuperNovae (ASAS-SN) catalogue \citep[][]{Jayasinghe2019}, which reports an amplitude of 0.45 mag and a period of 399 days. The latter is not the rotational period, but it is rather related to long-term variations of the starspot distribution. 
%The AAVSO International Variable Star Index VSX catalog \citep{Watson2006} 
The catalog of All Sky Automated Survey (ASAS) photometry of ROSAT sources-II \citep{Kiraga2013} lists a variation amplitude of 0.15 mag and a more reliable period of 13.63 days, which is close to the orbital one measured by us.
The orbital solution indicates a very low eccentricity that, according to \citet{Lucy1971}, can be considered as zero.

The timescales for circularization and synchronization, 
calculated for the primary component according to \citet{Zahn1989}, are $\tau_{\rm circ}\sim 3$\,Myr and $\tau_{\rm sync}\sim 0.05$\,Myr, respectively. This is in line with the observations that indicate a synchronous system with a nearly circular orbit.
The spectral subtraction reveals H$\alpha$ emission only from
(or predominantly from) the cooler, more massive component. This is not surprising, considering the larger stellar radius that we estimate for this star ($\sim$5\Rsun) compared to the secondary component ($\sim$2.5\Rsun), which implies a larger equatorial velocity for the primary component, as also indicated by the values of the projected rotational velocity (see Table\,\ref{Tab:phys}). This is producing a stronger dynamo action on sub-photospheric layers of the primary component, which is responsible for the higher level of magnetic activity, similarly to what is observed in well-known RS CVn systems like HR\,1099, WW\,Dra, and UX\,Ari \citep[e.g.,][]{Frasca1994,Montes1995}, whose components have similar SpT as TYC\,3386-868-1.

\subsection{G\,137-52}
In addition to X-ray emission, this star is also an extreme ultraviolet %(EUV) 
source included in the 2RE Source Catalogue \citep[][]{Pye1995} with the name
2RE\,J1547+150. It was already discovered as an SB2 by \citet{Jeffries1995} who obtained intermediate-dispersion spectra in a spectral range around the H$\alpha$ line. They also present a poorly sampled radial velocity curve (also due to the orbital period almost equal to 5 days) with relatively large errors (2--4\,\kms) and a solution with orbital parameters close to those found by us. In their spectra, the H$\alpha$ line displays a filled-in profile, but, without a subtraction of a photospheric template, they cannot say anything about the chromospheric activity level of both components.
From our spectra the secondary component displays a stronger filling of the H$\alpha$ line which, in some spectra, is just above the local continuum. 

\begin{figure}
\begin{center}
\hspace{-.5cm}
\includegraphics[width=8.5cm]{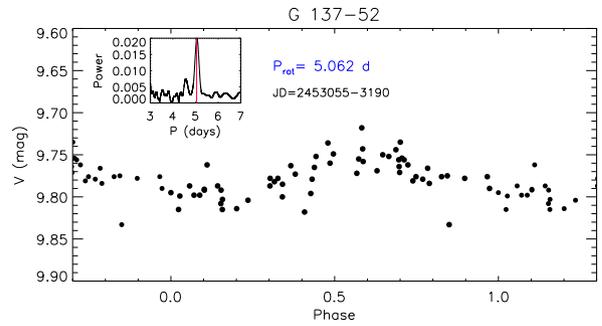}	
\vspace{0cm}
\caption{Phased $V$-band ASAS light curve of G\,137-52 from 19 February to 3 July 2004. The inset shows the cleaned periodogram of these data; the
orbital period is marked with a vertical red line.}
\label{fig:ASAS_GD137}
\end{center}
\end{figure}

No information on photometric period can be found in the literature, by seeking in the Vizier database. We thus searched for periods in the ASAS photometry available for this source  with the cleaned periodogram analysis. We found periods ranging from 4.99 to 5.13 days by analyzing different data segments with length shorter than 300 days (see Fig.\,\ref{fig:ASAS_GD137}). A peak at $P_{\rm rot}=5.01$\,days is clearly visible in the whole time series.
Therefore, G\,137-52 appears to be a synchronous system in a nearly circular orbit.
This is in line with the timescales for circularization and synchronization, which are $\tau_{\rm circ}\sim 5-10$\,Gyr and $\tau_{\rm sync}\sim 15$\,Myr, respectively.

\subsection{BD+10\,2953}
\begin{figure}
\begin{center}
\hspace{-.5cm}
\includegraphics[width=8.5cm]{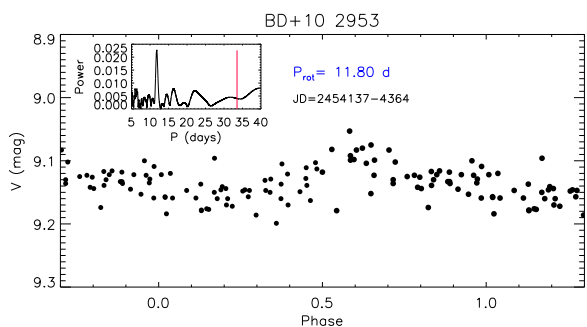}	
\vspace{0cm}
\caption{$V$-band ASAS light curve of BD+10\,2953 from 5 February to 21 September 2007 phased with the photometric period of 11.80 days. The inset shows the cleaned periodogram of these data; the
orbital period is marked with a vertical red line.}
\label{fig:ASAS_BD10}
\end{center}
\end{figure}

This object is classified as a variable source in the ASAS photometry of ROSAT sources-I catalogue \citep{Kiraga2012}, which reports a period of 12.19 days. 
The ASAS-SN catalogue \citep[][]{Jayasinghe2019} reports instead an amplitude of 0.15 mag and a period of 427 days, which is likely the result of long-term variations.
Indeed, if  we do a period search in portions of ASAS-SN V data spanning less than one year, we find cleaned periodograms without strong and clear peaks. Some indication of a period of about 27.9 days emerges for the data in the Julian day range JD=[2457020,2457250] and 30.3 days for JD=[2457397,2457630], respectively. However, in both cases a rather low peak amplitude of $\sim$\,0.025 mag was found.
%at  29.9 d and 27.9 d for the Julian day ranges [2456700,2456910] and [2457020,2457250], respectively. 
These periodicities are not far from the orbital period and might be related to the rotational modulation produced by starspots in one or both components or to proximity effects.
A similar analysis applied to segments of the ASAS photometry provided results in agreement with those of \citep{Kiraga2012}, since we found peaks of the cleaned periodograms in the range 11.6--12.4 days in different segments (see Fig.\,\ref{fig:ASAS_BD10}) and $P=11.8$\,days, as the second highest peak, in the full dataset.

The timescales for circularization and synchronization, 
calculated for the primary component according to \citet{Zahn1989}, are $\tau_{\rm circ}\sim 2.3$\,Gyr and $\tau_{\rm sync}\sim 10$\,Myr, respectively. For the secondary component a longer synchronization time, $\tau_{\rm sync}\sim 100$\,Myr, is found. As we did not detect the Li{\sc i}\,$\lambda$6708 absorption line from either component, the system should be much older than 10\,Myr and even older than 100\,Myr, which is close to the age of the Pleiades; therefore the system should have already attained the spin-orbit synchronization while the orbit is still highly eccentric.
However, in an eccentric orbit the tidal interaction is stronger at periastron, when the orbital velocity is higher, 
with the consequence that the equilibrium is reached at a value of rotation period, $P_{\rm pseudo}$, which is smaller than $P_{\rm orb}$, leading to a pseudo-synchronization \citep[see, e.g.,][]{Hut1981}. The value of $P_{\rm pseudo}$ depends on the orbital period and the eccentricity of the system and, following the guidelines of \citet{Hut1981}, results to be about 11.5 days for the components of BD+10\,2953.
%, which is similar to the period of 12.19 days reported by \citet{Kiraga2012}. 
The timescale for the pseudo-synchronization can be evaluated as $\tau_{\rm pseudo}\sim 23$\,Myr, which, according to the above arguments, is smaller than the system age.
The pseudo-synchronization period, $P_{\rm pseudo}\simeq 11.5$\,days, is close to the period derived  by \citet{Kiraga2012} and by our analysis of the ASAS photometry ($P$=11.6--12.4 days), but it has not been found in the  ASAS-SN data. 
More precise photometry, like that one collected by space missions, will be crucial to confirm or reject the pseudo-synchronization status of this system.  

The spectral subtraction reveals a moderate filling of the H$\alpha$ core of the primary component only.

\subsection{V1079~Her}
This star is included in the Hamburg/RASS Catalogue of optical identifications \citep{Zickgraf2003}, but the saturated prism-objective spectrum did not allow them to classify this object.
Various period determinations can be found in the literature for this source, since from its discovery as a variable star by \citet{Robb2003}, who report, from $VRI$ photometry, a period of 19.1 days and suggest a spotted late-type giant. 
The ASAS catalogue of variable stars \citep{Pojmanski2002} reports a period of 18.97 days and an amplitude of 0.12 mag.
\citet{Norton2007} quote a period of 18.5948 days from SuperWASP (Wide Angle Search for Planets) photometry.
We have reanalysed the ASAS data finding a peak at 19.04 days.
All these determinations are very close to the orbital period indicating that the spin-orbit synchronization has been attained in this system.
Indeed, the timescales for circularization and synchronization are  $\tau_{\rm circ}\sim 20$\,Myr and $\tau_{\rm sync}\sim 0.2$\,Myr, respectively.

The two components display also a comparable level of chromospheric activity as suggested by the similar intensity of the H$\alpha$ emission filling the cores of the profiles of the two components (Fig.\,\ref{fig:subtraction_halpha}).

\subsection{BD+62\,1880}
\begin{figure}
\begin{center}
\hspace{-.5cm}
\includegraphics[width=8.5cm]{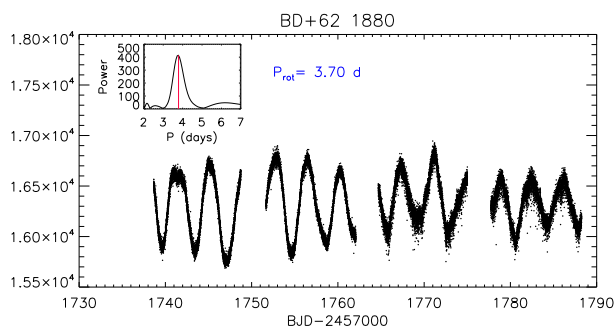}	
\vspace{0cm}
\caption{TESS light curve of BD+62\,1880 in 2019. The inset shows the cleaned periodogram of these data; the
orbital period is marked with a vertical red line.}
\label{fig:TESS_BD+62}
\end{center}
\end{figure}

No information on photometric period can be found in the literature, by seeking in the Vizier database of catalogs.
Fortunately, space-born accurate photometry was obtained with NASA\u2019s
 Transiting Exoplanet Survey Satellite \citep[TESS][]{Ricker2015}.
This object was observed in sector 16 between 2019-09-12 and 2019-10-06, in sector 17 between 2019-10-08 and 2019-11-02, and in sector 24 between 2020-04-16 and 2020-05-12. 
%The single aperture photometry (SAP) of the first two nearly  is plotted in
The first two nearly consecutive datasets are plotted in Fig.\,\ref{fig:TESS_BD+62} as a function of the barycentric Julian date (BJD) and display a clear rotational modulation. 
The cleaned periodogram, shown in the inset plot, displays a peak at 3.70\,days, which is close to the orbital period of the system.
The short timescale for synchronization, $\tau_{\rm sync}\sim 4$\,Myr, supports the results of the analysis of the TESS light curve.

The spectral subtraction (Fig.\,\ref{fig:subtraction_halpha}) shows H$\alpha$ emission from the secondary K0V component, while the primary
F9V component does not display any significant filling in the core of the line. 

\subsection{TYC\,4279-1821-1}

\begin{figure}
\begin{center}
\hspace{-.5cm}
\includegraphics[width=8.5cm]{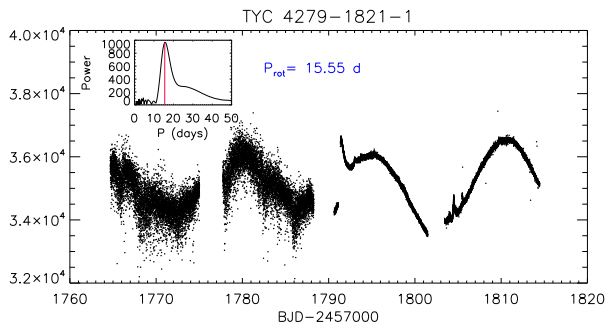}	
\vspace{0cm}
\caption{TESS light curve of TYC\,4279-1821-1 in 2019. The inset shows the cleaned periodogram of these data; the
orbital period is marked with a vertical red line.}
\label{fig:TESS_TYC4279}
\end{center}
\end{figure}

The first optical identification of this X-ray source was made by \citet{Motch1998}, who report a G0V SpT.
It is located in the field of the open cluster NGC\,7654, but neither its parallax nor its proper motions (see Table\,\ref{Tab:X}) are consistent with the average values for NGC\,7654, which are $\pi=0.596\pm 0.002$\,mas, $\mu_{\alpha}=-1.938\pm 0.005$\,mas/yr, and $\mu_{\delta}=-1.131\pm 0.005$\,mas/yr \citep{Cantat2018}.
It was classified as an eclipsing binary with a period $P=30.976$\,days by \citet{Laur2017}, but their light curve looks more like  a rotational  modulation with an amplitude of about 0.10 mag. It is worth noticing that this period is about twice the orbital one found by us and listed in Table\,\ref{Tab:RV}.
A rotational period of 15.618\,days is instead reported by \citet{Watson2006}.

TYC\,4279-1821-1 was observed by TESS in sector 17 between 2019-10-08, in sector 18 between 2019-11-03 and 2019-11-27, and in sector 24 between 2020-04-16 and 2020-05-12. The first two nearly consecutive datasets are plotted in Fig.\,\ref{fig:TESS_TYC4279} as a function of the BJD and display a clear rotational modulation with a period of 15.55\,days, which is exactly the orbital period, as also apparent from the peak of the periodogram.
A few flares are visible in the more precise TESS light curve taken in sector 18. 
Therefore, this system appears to be synchronous and with a circular orbit, in line with the circularization and synchronization times, $\tau_{\rm circ}\sim 5$\,Myr and $\tau_{\rm sync}\sim 0.1$\,Myr.

As for TYC\,3386-868-1 and other RS\,CVn systems, the H$\alpha$ emission is related to the cooler primary K0IV component, while the G5IV secondary does not display any significant filling of the line core.

As mentioned in Section~\ref{Sec:chrom_lithium}, this is the only system for which we could detect Li{\sc i}$\lambda$6708 absorption from both components. The equivalent width of the lithium line, corrected for the contribution to the continuum, is $EW_{\rm Li}^{\rm P}=140$\,m\AA\ and $EW_{\rm Li}^{\rm S}=280$\,m\AA, for the primary and secondary component, respectively, which give rise to lithium abundances of 2.2 and 3.1 dex.

\section{Conclusions}

In this work we have studied six spectroscopic binaries with X-ray emission, out of which five were recently discovered. We performed a high-resolution spectroscopic monitoring of our targets 
(TYC\,3386-868-1, G\,137-52, BD+10\,2953, V1079\,Her, BD+62\,1880, and, TYC\,4279-1821-1), from 2015 to 2021, mainly with the CAOS spectropolarimeter. As a result, we obtained the radial velocity curve for each binary, and from its analysis we determined the orbital parameters of the pair. Additionally, we estimated the spectral type and atmospheric parameters for both components of the system. From the comparison of the dynamical masses (inferred from the orbital solution) and the evolutionary ones (estimated from the location of the stars in the HR diagram) we derived the inclination of the system.

We find that our targets correspond to low-mass stars with masses between 0.8 and 1.5\,M$_{\odot}$ and spectral types from F\,9 to K\,4. G\,137-52 and BD+62\,188 are composed of main sequence stars while in the remaining four systems are involved evolved stars, similar to that observed in binaries of the RS\,CVn type. For all observed stars, luminosity classes agree with the evolutionary stages, as evidenced by their positions in the HR diagram.  
The orbital periods of the MS binaries are the shortest (3.7--5.0 d), whereas for the four binaries with evolved components they are larger (13-33 d).  
All systems studied in this work follow nearly circular orbits with the exception of BD+10\,2953, the system with the longest period, which exhibits a high eccentricity.

We also resorted to archival photometric data to search for photometric periods by means of periodogram analysis. For a few objects there were also period determinations in the literature.
We found that the five systems with a nearly circular orbit have also photometric periods close or equal to the orbital ones, which indicates that these system have already attained spin-orbit synchronization. This is in line with the timescales for synchronization, which have been evaluated according to the prescription of \citet{Zahn1989} and result to be always shorter than about 15\,Myr, i.e. much less than the age of these systems, as inferred, e.g., from the atmospheric lithium content.   
The longest circularization times are found for G\,137-52 ($\tau_{\rm circ}\geqsim 5$\,Myr) and BD+10\,2953 ($\tau_{\rm circ}\geqsim 2.3$\,Myr). 
These systems are also those with the highest eccentricities.
For BD+10\,2953, which displays the most eccentric orbit ($e=0.510$), a possible pseudo-synchronization with the periastron velocity is suggested.

Finally, we have also investigated the chromospheric activity finding that for G\,137-52 and V\,1079\,Her both components are active. In the remaining four systems the H$\alpha$ emission is only visible in the cooler component. Additionally, none of the pairs show an appreciable amount of photospheric Li, with the exception of TYC\,4279-1821-1, which exhibits high abundances in both components. %, likely due to the spin-orbit synchronization.  

\section*{Acknowledgments}
%We are grateful to the anonymous referee for a careful reading of the manuscript and very useful suggestions that helped us to improve our work.
This research has made use of the SIMBAD database and VizieR catalogue access tool,
operated at CDS, Strasbourg, France.\\
This work has made use of data from the European Space Agency (ESA)
mission \gaia\ ({\tt https://www.cosmos.esa.int/gaia}), processed by
the {\it Gaia} Data Processing and Analysis Consortium (DPAC,
{\tt https://www.cosmos.esa.int/web/gaia/dpac/consortium}). Funding
for the DPAC has been provided by national institutions, in particular
the institutions participating in the {\it Gaia} Multilateral Agreement.\\
This paper includes data collected by the TESS mission which are publicly available from the Mikulski Archive for Space Telescopes (MAST). Funding for the TESS mis-
sion is provided by the NASA\u2019s Science Mission -Directorate.
Support from the Italian {\it Ministero dell'Istruzione, Universit\`a e Ricerca} (MIUR) is also acknowledged. 

\section*{Data Availability}
The spectroscopic data underlying this paper will be shared on reasonable request to the corresponding author.
TESS photometric data are available at {\tt https://archive.st sci.edu/}. 

\bibliographystyle{mnras}
\bibliography{bin_caos.bib}

\label{lastpage} 

\newpage

\begin{appendix}
\section{Tables with individual values of radial velocity}

\begin{table}
\caption{Heliocentric radial velocities of the two components of TYC\,3386-868-1.}
\begin{center}
\begin{tabular}{lrcrcl}
\hline
\noalign{\smallskip}
 HJD     &  RV$_{\rm P}$   &  $\sigma_{\rm RV_P}$ &  RV$_{\rm S}$   &  $\sigma_{\rm RV_S}$ & Instrument \\
(2\,450\,000+) &  \multicolumn{2}{c}{(\kms)} & \multicolumn{2}{c}{(\kms)} & \\
\hline
\noalign{\smallskip}
  4459.7215 &    2.34  &   1.50  &    53.24  &   0.61  & SARG \\ 
  7331.6297 &  -39.75  &   0.69  &    88.40  &   0.58  & CAOS \\ 
  7336.5926 &   58.62  &   1.43  &   -11.84  &   0.84  & CAOS \\ 
 {\it 9152.6506} & {\it 22.35} & {\it 6.62} & {\it 22.35} & {\it 6.62} & CAOS \\  
  9153.6690 &   -3.64  &   0.36  &    51.35  &   0.47  & CAOS \\  
  9154.6183 &  -23.57  &   0.27  &    74.60  &   0.31  & CAOS \\ 
  9211.4989 &  -35.42  &   0.84  &    90.80  &   0.45  & CAOS \\ 
  9212.4992 &  -29.48  &   0.99  &    82.79  &   0.51  & CAOS \\ 
  9241.4697 &   -4.61  &   2.32  &    54.96  &   1.55  & CAOS \\ 
 {\it 9242.4122} & {\it 23.16} & {\it 1.64} & {\it 23.16} & {\it 1.64} & CAOS \\ 
  9250.4128 &   -1.27  &   3.42  &    52.76  &   2.05  & CAOS \\ 
  9251.3775 &  -21.33  &   0.89  &    74.95  &   0.52  & CAOS \\ 
 {\it 9263.3402} & {\it 23.80} & {\it 0.47} & {\it 23.80} & {\it 0.47} & CAOS \\	  
  9271.3725 &   56.50  &   1.84  &   -10.86  &   1.15  & CAOS \\ 
  9272.3788 &   78.70  &   1.97  &   -29.85  &   1.04  & CAOS \\ 
\noalign{\smallskip}
\hline
\end{tabular}
\end{center}
{\bf Notes.} RVs obtained close to the conjunctions are the same for the primary and secondary component and are written in italic.
%\begin{list}{}{}
%\item[$^a$] 
%\item[$^b$] 
%\end{list}
\label{Tab:RV_TYC3386}
\end{table}

\begin{table}
\caption{Heliocentric radial velocities of the two components of G\,137-52.}
\begin{center}
\begin{tabular}{lrcrcl}
\hline
\noalign{\smallskip}
 HJD     &  RV$_{\rm P}$   &  $\sigma_{\rm RV_P}$ &  RV$_{\rm S}$   &  $\sigma_{\rm RV_S}$ & Instrument \\
(2\,450\,000+) &  \multicolumn{2}{c}{(\kms)} & \multicolumn{2}{c}{(\kms)} & \\
\hline
\noalign{\smallskip}
  2439.3928  &   57.43  &   1.29  &  -10.19  &   1.48 & AURELIE \\
  2441.3860  &   -7.34  &   1.43  &   54.13  &   1.36 & AURELIE \\ 
  6749.6221  &   37.42  &   0.13  &    7.79  &   0.29 & CAOS \\   
  6805.4164  &   46.08  &   0.16  &   -2.18  &   0.28 & CAOS \\   
  6818.3565  &   -9.41  &   0.11  &   56.40  &   0.21 & CAOS \\   
  6819.4305  &    3.11  &   0.11  &   41.73  &   0.19 & CAOS \\   
  6839.3265  &   -4.75  &   0.10  &   52.41  &   0.19 & CAOS \\   
  6845.4400  &   34.35  &   0.25  &   11.23  &   0.40 & CAOS \\   
  6846.3591  &   56.45  &   0.11  &  -12.39  &   0.20 & CAOS \\   
  7204.3486  &   49.24  &   0.12  &   -5.73  &   0.23 & CAOS \\   
  7262.2984  &   -2.08  &   0.13  &   50.05  &   0.24 & CAOS \\   
  7263.3128  &   -6.80  &   0.12  &   53.78  &   0.22 & CAOS \\   
  7892.5272  &   30.57  &   0.20  &   14.61  &   0.39 & CAOS \\   
  7897.4103  &   36.94  &   0.13  &    8.45  &   0.24 & CAOS \\   
  7919.4674  &   -9.35  &   0.14  &   56.24  &   0.23 & CAOS \\   
  {\it 7920.4305}  & {\it 22.68} & {\it 0.05} & {\it 22.68} & {\it 0.05} & CAOS \\   
  7934.4536  &  -11.01  &   0.12  &   58.19  &   0.19 & CAOS \\   
  7941.4088  &   49.82  &   0.12  &   -4.96  &   0.19 & CAOS \\  
\noalign{\smallskip}
\hline
\end{tabular}
\end{center}
{\bf Notes.} RVs obtained close to the conjunctions are the same for the primary and secondary component and are written in italic.
\label{Tab:RV_G137-52}
\end{table}

\begin{table}
\caption{Heliocentric radial velocities of the two components of BD+10\,2953.}
\begin{center}
\begin{tabular}{lrcrcl}
\hline
\noalign{\smallskip}
 HJD     &  RV$_{\rm P}$   &  $\sigma_{\rm RV_P}$ &  RV$_{\rm S}$   &  $\sigma_{\rm RV_S}$ & Instrument \\
(2\,450\,000+) &  \multicolumn{2}{c}{(\kms)} & \multicolumn{2}{c}{(\kms)} & \\
\hline
\noalign{\smallskip}
  4148.6430  &  -50.91  &  0.80  &   -15.19  &  0.92 & SARG \\
  6729.6474  &  -53.57  &  0.09  &   -16.48  &  0.20 & CAOS \\
  6734.6523  &  -42.17  &  1.82  &   -31.85  &  1.80 & CAOS \\     
  6776.5294  &  -22.06  &  0.09  &   -50.40  &  0.20 & CAOS \\   
  6818.4420  &   -4.51  &  0.10  &   -71.42  &  0.20 & CAOS \\   
  6820.4257  &  -26.89  &  0.14  &   -48.72  &  0.25 & CAOS \\   
  6845.4724  &  -18.15  &  0.04  &   -57.53  &  0.06 & CAOS \\   
  6846.4040  &  -15.59  &  0.10  &   -60.02  &  0.20 & CAOS \\   
  6852.3726  &   -8.46  &  0.12  &   -70.88  &  0.21 & CAOS \\   
  7178.5403  &  -23.14  &  1.00  &   -51.75  &  0.94 & CAOS \\   
  {\it 7223.3655}  & {\it -35.82} & {\it 0.06} & {\it -35.82} & {\it 0.06} & CAOS \\   
  7263.2980  &  -60.76  &  0.17  &    -8.72  &  0.26 & CAOS \\   
  7568.4804  &  -53.22  &  0.10  &   -20.10  &  0.19 & CAOS \\   
  7892.5101  &  -14.11  &  0.11  &   -61.11  &  0.20 & CAOS \\   
  7897.3957  &  -67.88  &  0.10  &    -1.51  &  0.20 & CAOS \\   
  7919.3978  &  -14.17  &  0.10  &   -59.83  &  0.19 & CAOS \\   
  7920.3761  &  -11.70  &  0.10  &   -63.26  &  0.17 & CAOS \\   
  7933.4627  &  -61.61  &  0.10  &    -6.81  &  0.18 & CAOS \\  
  7941.4877  &  -41.82  &  0.76  &   -29.85  &  1.19 & CAOS \\     
  {\it 8262.4422}  & {\it -35.52} & {\it 0.05} & {\it -35.52} & {\it 0.05} & CAOS \\
  8277.4141  &  -39.84  &  0.40  &   -32.33  &  0.61 & CAOS \\
  8290.4077  &   -7.73  &  0.10  &   -66.55  &  0.19 & CAOS \\
  8299.4042  &  -68.25  &  0.11  &    -1.93  &  0.19 & CAOS \\
  8305.4078  &  -52.87  &  0.13  &   -16.38  &  0.30 & CAOS \\
  {\it 8312.3615}  & {\it -38.00} & {\it 0.05} & {\it -38.00} & {\it 0.05} & CAOS \\
  9300.5913  &  -14.47  &  0.26  &   -59.43  &  0.25 & CAOS \\ 
\noalign{\smallskip}
\hline
\end{tabular}
\end{center}
{\bf Notes.} RVs obtained close to the conjunctions are the same for the primary and secondary component and are written in italic.
\label{Tab:RV_BD+10_2953}
\end{table}

\begin{table}
\caption{Heliocentric radial velocities of the two components of V1079~Her.}
\begin{center}
\begin{tabular}{lrcrcl}
\hline
\noalign{\smallskip}
 HJD     &  RV$_{\rm P}$   &  $\sigma_{\rm RV_P}$ &  RV$_{\rm S}$   &  $\sigma_{\rm RV_S}$ & Instrument \\
(2\,450\,000+) &  \multicolumn{2}{c}{(\kms)} & \multicolumn{2}{c}{(\kms)} & \\
\hline
\noalign{\smallskip}
  2439.4846  &   17.91  &   1.57  &  -45.20  &  1.38 & AURELIE \\
  2441.5015  &   41.48  &   1.52  &  -69.37  &  1.31 & AURELIE \\ 
  2443.3980  &   40.14  &   1.40  &  -69.29  &  1.30 & AURELIE \\   
  6776.5673  &  -40.38  &   0.28  &    9.18  &  0.51 & CAOS \\     
  6818.5406  &  -74.12  &   0.21  &   42.66  &  0.38 & CAOS \\     
  6839.4504  &  -68.94  &   0.64  &   36.57  &  1.23 & CAOS \\     
  6845.5206  &   23.98  &   0.31  &  -54.71  &  0.43 & CAOS \\     
  6846.4786  &   34.78  &   0.41  &  -67.58  &  0.51 & CAOS \\     
  6852.3969  &   -4.24  &   3.94  &  -23.04  &  2.59 & CAOS \\   
  8262.5936  &   27.25  &   0.20  &  -58.17  &  0.37 & CAOS \\     
  8277.4454  &  -49.05  &   0.18  &   17.99  &  0.29 & CAOS \\     
  8291.4044  &  -55.76  &   0.20  &   21.52  &  0.34 & CAOS \\     
  8299.4344  &   -5.57  &   0.26  &  -26.37  &  0.53 & CAOS \\     
  8311.4260  &  -64.69  &   0.21  &   29.02  &  0.36 & CAOS \\     
  8319.4545  &    5.23  &   0.27  &  -40.88  &  0.55 & CAOS \\     
  8330.3594  &  -58.57  &   0.22  &   22.48  &  0.34 & CAOS \\
\noalign{\smallskip}
\hline
\end{tabular}
\end{center}
\label{Tab:RV_V1079Her}
\end{table}

\begin{table}
\caption{Heliocentric radial velocities of the two components of BD+62\,1880.}
\begin{center}
\begin{tabular}{lrcrcl}
\hline
\noalign{\smallskip}
 HJD     &  RV$_{\rm P}$   &  $\sigma_{\rm RV_P}$ &  RV$_{\rm S}$   &  $\sigma_{\rm RV_S}$ & Instrument \\
(2\,450\,000+) &  \multicolumn{2}{c}{(\kms)} & \multicolumn{2}{c}{(\kms)} & \\
\hline
\noalign{\smallskip}
   7261.5032 &   31.50 &  0.22 &  -77.29  &  0.79 & CAOS \\ 
   7263.4735 &  -53.11 &  0.21 &   38.93  &  0.76 & CAOS \\ 
   7308.3951 &  -82.11 &  0.18 &   80.32  &  0.66 & CAOS \\  
   7336.4194 &   38.35 &  0.17 &  -86.03  &  0.61 & CAOS \\  
   7599.5556 &  -38.61 &  0.16 &   20.06  &  0.53 & CAOS \\ 
   8306.5822 &   13.35 &  0.19 &  -59.15  &  0.72 & CAOS \\ 
   8311.5974 &   25.09 &  0.15 &  -72.08  &  0.54 & CAOS \\ 
   8312.5622 &  -68.60 &  0.18 &   62.09  &  0.66 & CAOS \\ 
   8330.5611 &   23.23 &  0.20 &  -71.17  &  0.82 & CAOS \\ 
   8367.4383 &   34.56 &  0.18 &  -78.81  &  0.71 & CAOS \\ 
   9154.2913 &  -77.93 &  0.22 &   75.53  &  0.59 & CAOS \\ 
   9446.4844 &  -79.72 &  0.53 &   75.82  &  0.99 & CAOS \\ 
   9453.4881 &  -60.65 &  0.70 &   55.32  &  0.98 & CAOS \\ 
   9454.4461 &  -59.39 &  0.63 &   50.90  &  1.15 & CAOS \\ 
   9455.4301 &   37.36 &  0.54 &  -86.94  &  0.90 & CAOS \\ 
   9459.4709 &   52.30 &  0.89 & -101.16  &  1.21 & CAOS \\
   9467.4351 &   50.87 &  0.72 &  -96.84  &  1.14 & CAOS \\
\noalign{\smallskip}
\hline
\end{tabular}
\end{center}
\label{Tab:RV_BD+62_1880}
\end{table}

\begin{table}
\caption{Heliocentric radial velocities of the two components of TYC\,4279-1821-1.}
\begin{center}
\begin{tabular}{lrcrcl}
\hline
\noalign{\smallskip}
 HJD     &  RV$_{\rm P}$   &  $\sigma_{\rm RV_P}$ &  RV$_{\rm S}$   &  $\sigma_{\rm RV_S}$ & Instrument \\
(2\,450\,000+) &  \multicolumn{2}{c}{(\kms)} & \multicolumn{2}{c}{(\kms)} & \\
\hline
\noalign{\smallskip}
  2230.3281   &  33.13  & 1.49  & -53.17 &  1.53 & AURELIE \\
  2234.3389   & -19.19  & 1.59  &   4.75 &  1.80 & AURELIE \\ 
  7262.5687   & -47.94  & 0.15  &  35.93 &  0.25 & CAOS \\	
  7331.5103   &  30.93  & 0.14  & -54.42 &  0.20 & CAOS \\	
  7336.4839   & -19.95  & 0.15  &   4.35 &  0.27 & CAOS \\	
  9152.3811   &  30.50  & 0.16  & -54.15 &  0.22 & CAOS \\	  
  9152.4050   &  30.22  & 0.20  & -53.72 &  0.24 & CAOS \\    
  9154.4664   &  10.63  & 0.23  & -32.32 &  0.37 & CAOS \\	
  9223.3211   & -43.73  & 2.36  &  28.54 &  1.86 & CAOS \\   
  9242.2367   &   5.95  & 2.45  & -25.30 &  1.65 & CAOS \\	
  9250.2387   & -26.04  & 1.12  &  11.04 &  1.02 & CAOS \\    
  9251.2569   & -39.48  & 0.59  &  26.42 &  0.48 & CAOS \\    
  9453.5268   & -39.84  & 0.36  &  28.24 &  0.31 & CAOS \\     
\noalign{\smallskip}
\hline
\end{tabular}
\end{center}
\label{Tab:RV_TYC4279}
\end{table}

\end{appendix}

\end{document}